\documentclass[fleqn,usenatbib]{mnras}

\usepackage{newtxtext,newtxmath}

\usepackage[T1]{fontenc}

\DeclareRobustCommand{\VAN}[3]{#2}
\let\VANthebibliography\thebibliography
\def\thebibliography{\DeclareRobustCommand{\VAN}[3]{##3}\VANthebibliography}

\usepackage{graphicx}	
\usepackage{amsmath}	

\usepackage{xcolor}

\newcommand{\rvs}[1]{#1}
\newcommand{\raul}[1]{#1}

\title[Constraints on Baryon Feedback]{Determining the Baryon Impact on the Matter Power Spectrum with Galaxy Clusters}

\author[SG et al.]{
Sebastian Grandis,$^{1}$\thanks{E-mail: sebastian.grandis@uibk.ac.at}
Giovanni Aric\`o,$^{2}$
Aurel Schneider,$^{2}$
Laila Linke$^{1}$
\\
$^{1}$Universit\"at Innsbruck, Institut f\"ur Astro- und Teilchenphysik, Technikerstr. 25/8, 6020 Innsbruck, Austria\\
$^{2}$Institute for Computational Science, University of Zurich, Winterthurerstrasse 190, 8057 Zurich, Switzerland\\
}

\date{Accepted XXX. Received YYY; in original form ZZZ}

\pubyear{2015}

\begin{document}
\label{firstpage}
\pagerange{\pageref{firstpage}--\pageref{lastpage}}
\maketitle

\begin{abstract}
The redistribution of baryonic matter in massive halos through processes like active galactic nuclei feedback and star formation leads to a suppression of the matter power spectrum on small scales. This redistribution can be measured empirically via the gas and stellar mass fractions in galaxy clusters, and leaves imprints on their electron density profiles. We constrain two semi-analytical baryon correction models with a compilation of recent Bayesian population studies of galaxy groups and clusters sampling a mass range above $\sim 3 \times 10^{13}$  $M_\odot$, and with cluster gas density profiles derived from deep, high-resolution X-ray observations. We are able to fit all the considered observational data, but highlight some anomalies in the observations.
The constraints allow us to place precise, physically informed priors on the matter power spectrum suppression. At a scale of $k=1 h$ Mpc$^{-1}$ we find a suppression of $0.042^{+0.012}_{-0.014}$ ($0.049^{+0.016}_{-0.012}$), while at $k=3h$ Mpc$^{-1}$ we find $0.184^{+0.026}_{-0.031}$
($0.179^{+0.018}_{-0.020}$), depending on the model used. In our fiducial setting, we also predict at 97.5 percent credibility, that at scales $k<0.37h$ Mpc$^{-1}$ baryon feedback impacts the matter power less than $1\%$. This puts into question if baryon feedback is the driving factor for the discrepancy between cosmic shear and primary CMB results. We independently confirm results on this suppression from small-scale cosmic shear studies, while we exclude some hydro-dynamical simulations with too strong and too weak baryonic feedback. Our empirical prediction of the power spectrum suppression shows that studies of galaxy groups and clusters will be instrumental in unlocking the cosmological constraining power of future cosmic shear experiments like \textit{Euclid} and Rubin-LSST, and invites further investigation of the baryon correction models.
\end{abstract}




\section{Introduction}

Cosmological inference on the matter distribution of the Universe from future surveys like \textit{Euclid}\footnote{\url{http://www.euclid-ec.org/}}, LSST\footnote{\url{https://www.lsst.org/}}, or Roman\footnote{\url{https://roman.gsfc.nasa.gov/}} will be limited by our knowledge of non-gravitational effects on the matter power spectrum \citep[for reviews, see][and reference therein]{chisari19, Eckert21}. Specifically, on physical scales below $\sim10$ Mpc, active galactic nuclei are known to lead to a significant redistribution of a fraction of the baryonic matter, while on even smaller scales star formation allows another fraction of baryons to condensate into galaxies inhabiting the halo centers. Both these redistribution mechanisms lead to a gravitational back reaction on the collisionless dark matter, further altering the matter distribution. These effects are collectively referred to as $\textit{baryon feedback}$, and are degenerate with interesting cosmological signals like neutrino masses, dark energy equation of state modifications, modified gravity signatures or non cold dark matter components \citep[e.g.][]{Harnois15, schneider20}.

While hydro-dynamical simulations are in principle able to accurately predict baryon feedback effects on cosmology, practical concerns like run-time and resolution limits require the tuning of several recipes that attempt the modelling of baryon feedback by summarizing sub resolution processes \citep{schaye15,maccarthy17, pillepich18}. While providing physically self-consistent solutions, hydro-dynamical simulations are thus only able to present point estimates for individual, best guess feedback models. They lack the predictive power and flexibility to be employed in cosmological inference tasks, as they probe only dozens of feedback models \citep{mead15, chisari18, vandaalen20}, though noticeable progress has been made by the recent CAMELS \citep{CAMELS:2020cof}, MillenniumTNG \citep{milleniumTNG}\footnote{\url{https://www.mtng-project.org/}},  ANTILLES \citep{Antilles2023}, and FLAMINGO\footnote{\url{https://flamingo.strw.leidenuniv.nl/}} \citep{flamingo1, flamingo2} projects. 

Filling this gap are so called Baryon Correction Models \citep[hereafter BCM, originally proposed by ][]{schneider15}. Starting from assumptions on the fraction and distribution of baryonic matter in and around halos, these models alter gravity-only simulations in a semi-analytical way, to mimic the impact of baryon feedback. Recently, \citet{arico20, arico21a} and \citet{giri21} have shown that their respective BCMs are able simultaneously to fit the stellar and gas components of halos over several orders of magnitude in mass, as well as the power and bispectrum in different hydro-dynamical simulations with a handful of physically motivated parameters. The resulting gas and stellar fractions as well as suppression of the matter power and bispectrum are easy to compute. This provides a significant benefit compared to feedback prescriptions that only parameterize the matter power spectrum suppression, as for instance presented by \citet{mead21}, and offer no predictive power on other quantities.

First attempts to constrain these BCMs have been undertaken with a variety of data sets.   \citet{schnieder21} used Kilo Degree Survey cosmic shear, Atacama Cosmology Telescope kinematic Sunyaev-Zeldovich profiles and a compilation of galaxy cluster and group measurements to simultaneously fit for the cosmological parameters and the BCM. On the cluster and group side that work was limited by the use of individual hydrostatic mass estimators \raul{from compilations of X-ray observed objects. Such compilations suffer from a lack of selection effects and mass calibration modelling, as they assume that the compiled list of clusters and groups is a fair sample of the underlying halo population, and that their halo masses can be calibrated by specifying on single parameter, the hydrostatic mass bias.} \citet{chen+22} and \citet{arico23} have derived first constraints on BCMs using Dark Energy Survey year 3 cosmic shear on small scale, and on all scales, respectively. Both works find weak constraints on one of the seven parameters of the BCM.

Parallel to these studies, major progress has been achieved in the observational study of galaxy clusters, which inhabit massive halos. Bayesian Population models simultaneously describe cluster selection and mass calibration, preferentially via the weak gravitational lensing (WL) signal these massive objects impress on background galaxies \citep{bocquet15, mantz15, sereno17, dietrich19, chiu22a}. The robustness of these modelling techniques is demonstrated by the fact that the number counts of clusters are an independent, competitive cosmological probe \citep{mantz15, bocquet19, chiu22cosmo}. This indicates that the mass distribution of these samples can be reliably reconstructed in the aforementioned Bayesian frameworks. Additionally, multi-wavelength studies of samples of galaxy clusters measure the gas and stellar mass in massive halos, and its mass trend \citep[see for instance][]{mantz16, chiu18, akino22, chiu22a}. Crucially, these studies forego the use of hydrostatic masses in favor of the more accurate WL signal. \raul{This lifts the uncertainties on sample selection and mass calibration that plague compilation of X-ray observed clusters, like the one recently presented by \citet{flamingo2}.} 

In this work, we combine recent weak lensing informed gas and stellar mass fraction measurements of massive halos \citep{mantz16, chiu18, akino22, chiu22a} to inform the BCM and predict the matter power spectrum suppression due to baryon feedback. The introduction of weak lensing mass calibration allows us to avoid using hydrostatic masses, significantly improving our mass accuracy.  We also include measurements of the gas profile from deep, high-resolution X-ray observations \citep{ghirardini19} to narrow down the shape of the gas density profiles of massive halos.

Halo masses in this work are reported as spherical over density masses $M_{\Delta \rm{c}}$, with $\Delta=500, 200$. This means that they are defined via the radius $R_{\Delta \rm{c}}$ enclosing a sphere of average density $\Delta$ times the critical density of the Universe at that redshift, i.e. $M_{\Delta \rm{c}} = \frac{4 \pi}{3} \Delta \rho_\text{crit}(z)R_{\Delta \rm{c}}^3$. Throughout this work, we use a flat $\Lambda$CDM cosmology with parameters $\Omega_\text{M}=0.315$, $\sigma_8 = 0.83$, $n_\text{S}=0.96$ and $h=0.67$ as reference cosmology.

\section{Data}\label{sec:data}

\begin{figure}
	\includegraphics[width=\columnwidth]{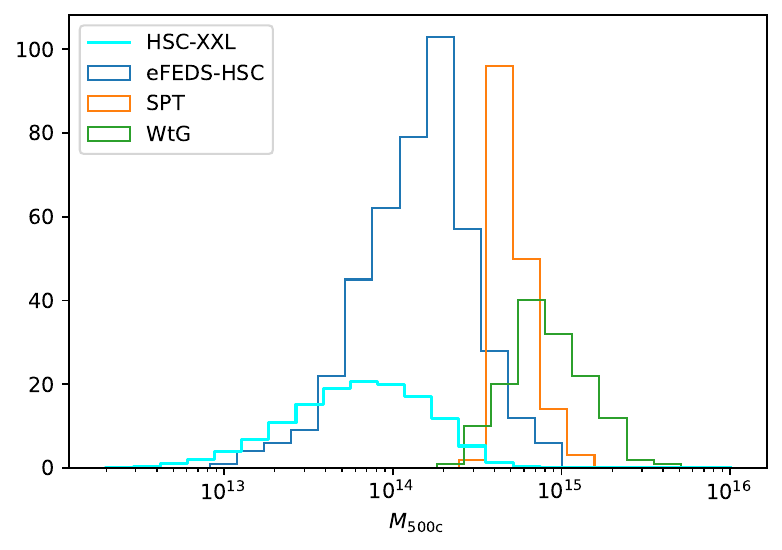}
    \caption{Approximate mass distributions of the galaxy cluster and group samples used. These distributions are not used in our inference and are presented for visualisation purposes only. Our compilation of studies samples the mass range $M_{500\text{c}} \gtrapprox 3 \times 10^{13}$  $M_\odot$. Sources: eFEDS-HSC \citet[][Tab.~C1]{chiu22a}; SPT \citet[with $\xi>6.8$]{bocquet19}; HSC-XXL redshifts from \citet[][Tab.~2]{umetsu20} together the mass distribution parameters from \citet{akino22}; WtG \citet[][Tab.~2 and Eq.~1]{mantz16}.}
    \label{fig:mass_distr}
\end{figure}

\begin{table}
\caption{\label{tab:fracs} Baryonic fractions and mass trends of weak lensing calibrated cluster and group population studies compiled by this work.}
\begin{tabular}{ccccc}
 & comp. &  $M_\text{piv}$ [$M_\odot$] & fraction  & slope \\ 
\hline
SPT & ICM & 4.80e+14 & $0.12 \pm 0.01$ & $0.33 \pm 0.09$ \\
SPT & stars & 4.80e+14 & $0.008 \pm 0.001$ & $-0.20 \pm 0.12$ \\
eFEDS & ICM & 2.00e+14 & $0.05 \pm 0.01$ & $0.19 \pm 0.11$ \\
HSC-XXL & ICM & 1.00e+14 & $0.08 \pm 0.01$ & $0.23 \pm 0.12$ \\
HSC-XXL & stars & 1.00e+14 & $0.021 \pm 0.003$ & $-0.20 \pm 0.11$ \\
WtG & ICM & 1.00e+15 & $0.12 \pm 0.00$ & \\
\end{tabular}
\end{table}

\subsection{Weak lensing calibrated mass fractions}

In this work we use measurements of the relation between the gas / stellar mass and the halo mass in galaxy clusters to constrain  Baryon Correction Models. We use the following compilation of stellar and gas mass fractions extracted from Bayesian Population models, summarized also in Table~\ref{tab:fracs}: 
\begin{itemize}
    \item The analysis of 139 ROSAT selected clusters, followed up with Chandra imaging analysed by \citet{mantz16}, 26 of which have dedicated weak lensing measurements. This analysis, called Weighing the Giants (WtG), reports a hot gas fraction $f_\text{ICM}= 0.125 \pm 0.005$ at pivot mass $M_\text{piv}=10^{15}$  $M_\odot$\footnote{Note that all these observational results measure baryonic fractions within $r_\text{500c}$, and w.r.t. halo masses $M_\text{500c}$.}. The mass information comes from a combination of cluster abundance measurements and weak lensing. We discard the slope measurement of this work, as it has implausibly tight error bars that would otherwise dominate our fitting.
    \item The study of 91 South Pole Telescope (SPT) selected clusters, whose Chandra, DES, Wise and Spitzer follow-up has been studied by \citet{chiu18}, reporting a hot gas fraction $f_\text{ICM}=0.119 \pm 0.013$ and a mass trend of the gas mass $B_\text{ICM} = 1.33 \pm 0.09$ at $M_\text{piv}=4.8\times10^{14}$  $M_\odot$, as well as stellar mass fraction $10^3 f_\star=8.3 \pm 0.6$ and mass trend of the stellar mass $B_\star = 0.80 \pm 0.13$. The mass information in this analysis comes from fitting the cluster abundance measurements \citep{dehaan+16}. The resulting mass calibration has been confirmed via weak lensing \citep{dietrich19, stern19, bocquet19, Schrabback21} and dynamical analysis \citep{capasso19}.
    \item The analysis of 136 galaxy clusters and groups by \citet{akino22}, which we call HSC-XXL, reporting $f_\text{ICM}=0.075 \pm 0.008$ and $B_\text{ICM} = 1.23 \pm 0.12$ at $M_\text{piv}=10^{14}$  $M_\odot$, as well as $10^3 f_\star=21.2 \pm 2.8$ and $B_\star = 0.80 \pm 0.11$. These objects were selected in the XMM-Newton XXL survey, and have HSC-SSP weak lensing measurements, XMM-Newton gas mass measurements and stellar mass measurements for HSC and SDSS photometry.  
    \item The study by \citet{chiu22a} of 434 clusters and groups selected in the eROSITA Final Equatorial Depth Survey (eFEDS), 313 of which have HSC-SSP weak lensing measurements, reporting $f_\text{ICM}=0.0540 \pm 0.0065$ and $B_\text{ICM}=1.19\pm 0.11$ at $M_\text{piv}=2\times 10^{14}$  $M_\odot$. 
\end{itemize}

The selection of these samples is performed via their extended Bremsstrahlung emission in the X-rays \rvs{(HSC-XXL, eFEDS, WtG)}, or via the Sunyaev-Zel'dovich effect (SZe, see \citet{Carlstrom02} for a review) in (sub-)millimeter observation \rvs{(SPT)}, i.e via inverse Compton scattering signatures in cosmic microwave background observations, and confirmed in optical wavelength. These methods provide high purity samples of massive halos, albeit with different redshift trends in the limiting mass. The information on the mass scale of these samples comes from weak lensing measurements in dedicated deep photometric follow-up \rvs{(WtG)} or deep photometric survey data \rvs{(HSC-XXL, eFEDS)}, and in some cases from abundance matching to the theoretical halo mass function \rvs{(SPT, WtG)}. Their gas mass is measured in high angular resolution X-ray follow-ups \rvs{(SPT, WtG)} or X-ray survey data \rvs{(HSC-XXL, eFEDS)}. The stellar masses are determined with optical and near-infrared observations of the cluster and group member galaxies. We refer the interested reader to the respective papers for more information.

Of interest to this work is that the different samples have minimal to no overlap in actual objects. We can thus treat the different measurements as mutually independent. Also, their mass distributions probe significantly different mass ranges, as seen in Fig.~\ref{fig:mass_distr}. These mass distributions are derived outcomes of the Bayesian Population analyses described below (cf. section~\ref{sec:bay_pop_mod}), and are not directly used in this analysis. We instead employ only the reconstructed observable halo mass relations.

\begin{table}
\caption{\label{tab:ne_prof} Mean electron density profile $n_e$, and its error  $\delta n_e$, as a function of radius, from the X-COP sample at median redshift  $z_\text{med}=0.064$ and median hydrostatic mass $M^\text{hydro}_{\text{med}, 500c}=3.79\times 10^{14}\, h^{-1} M_\odot$, adapted from \citet[][Tab.~2]{ghirardini19}.}
\begin{tabular}{ccc}
$r$ [Mpc$h^{-1}$] & $n_e$ [$10^{-3}$ cm$^{-3}$] & $\delta n_e$ [$10^{-3}$ cm$^{-3}$] \\ 
\hline
0.034 & 6.159 & 0.605 \\
0.085 & 3.371 & 0.380 \\
0.145 & 2.240 & 0.196 \\
0.221 & 1.395 & 0.085 \\
0.327 & 0.886 & 0.031 \\
0.502 & 0.445 & 0.011 \\
0.791 & 0.179 & 0.006 \\
1.339 & 0.053 & 0.004 \\
\end{tabular}
\end{table}

\subsection{Electron density profiles}

We complement the measurements of the relation between gas / stellar mass and halo mass for large populations of clusters with information on the gas density profile of 13 high mass, low redshift clusters observed with deep and high resolution dedicated X-ray imaging, carried out by the  XMM-Newton Cluster Outskirts Project \citep[X-COP, ][]{eckert17}. Analysis of their X-ray surface brightness and spectral information by \citet{ghirardini19} resulted in a measurement of the hydro-static mass of these objects, with median hydrostatic mass $M^\text{hydro}_{\text{med}, 500c}=3.79\times 10^{14}\, h^{-1} M_\odot$, as well as their electron density profile $n_e\left(r/r^\text{hydro}_{500c}\right)$. We utilize the results from the piece wise linear interpolation of the electron density profiles given in \citet[][Tab.~2]{ghirardini19} and the median redshift of the X-COP sample, $z_\text{med}=0.064$, to generate a measurement of the electron density profile $n_e$ in units of cm$^{-3}$ at 8 radii $r$ in units of Mpc$h^{-1}$. The resulting data vector is reported in Tab.~\ref{tab:ne_prof}. The resulting electron density profile is shown in Fig.~\ref{fig:ne_fitted} as black points. Note that we scaled the electron density profile by the radius $r$ to reduce the dynamical range in the plot. Many more high resolution X-ray measurements of the electron density of clusters would, in principle, be available \citep[e.g. ][]{croston08, mcdonald13, sanders17, bulbul19}. Adapting their results to our fitting procedure would, however, far exceed the scope of this work.

\section{Method}

We shall describe in the following how the multi-wavelength observations of galaxy groups and clusters are analysed to extract the gas and stellar mass fractions as a function of halo mass. We then describe how we use the BCM to fit these fractions, as well as the electron density profile of halos. Finally, we outline our fitting procedure and the adopted priors.

\subsection{Bayesian Population Models}\label{sec:bay_pop_mod}

The galaxy cluster and group studies used in this work, and listed in Section~\ref{sec:data}, infer the relation between the gas / stellar mass $M_{\text{ICM},\star}$ enclosed in $R_{500\text{c}}$ and the halo mass $M_{500\text{c}}$ via Bayesian Population models. These models are underpinned by a statistical prescription that describes the generation of the cluster catalog data starting from the parent distribution in mass and redshift space, $P(M, z)$ (often the halo mass function times the cosmological volume). A stochastical mapping from halo mass $M$ and redshift $z$ to intrinsic (noise free) observables $\mathcal{O}$ is then applied, $P(\mathcal{O}| M, z)$. Motivated by empirical evidence and simulation results \citep{angulo12}, this mapping is modelled as a multivariate log-normal distribution in the observables. The means are power laws in mass and redshift, called \emph{mass observable relations}. The simple power law relation is justified by the fact that clusters deviate only to second order from self similar behaviour characteristic of gravity and adiabatic thermodynamics\footnote{\citet[][sec. 3]{Mulroy19} presents an excellent review of the original derivation by \citet{Kaiser86}.}. The multivariate scatter around these mean relations is an expression of the heterogeneity of the cluster population at a given mass and redshift \citep[see, for instance, ][]{farahi19}. Measurement noise in the observables is modelled via a stochastical mapping between intrinsic (noise-free) observables and measured observables, $P(\hat{\mathcal{O}}| \mathcal{O}, z)$. This last mapping is determined directly from the noise properties of the observations, or via image simulations.

The distribution of observed cluster properties $P(\hat{\mathcal{O}}, z)$ is then obtained by marginalising of mass and intrinsic properties 

\begin{equation}
    P(\hat{\mathcal{O}}, z) = \int \text{d}M\, P(M, z) \int \text{d}\mathcal{O} P(\mathcal{O}| M, z) P(\hat{\mathcal{O}}| \mathcal{O}, z).
\end{equation}
This distribution is normalized to account for the selection criteria imposed in the sample's selection. The likelihood of a sample of clusters $\big\{ \hat{\mathcal{O}}_i, z_i \big\}$ results from evaluating the probability of this sample given the constructed distribution 

\begin{equation}
    \ln \mathcal{L} = \sum_i \ln P(\hat{\mathcal{O}}_i, z_i),
\end{equation}
as in Bayesian inference, the likelihood is simply the probability of the data (here the catalog) given the model (the distribution of observed cluster properties). This analysis approach ensures that selection biases and mass calibration uncertainties are correctly accounted for.

In the analyses we use for this work, the measured observables include the gas mass $\hat{M}_\text{ICM}$, the weak lensing signal in the form of a shear profile, or a best fit mass, and, where applicable, the stellar mass $\hat{M}_\star$, among other observables. For further details on each cluster study we refer the reader to the relative papers and references therein, noting that many details of the implementation, as well as the individual notations used to describe the Bayesian population model might differ significantly between the different works. Our notations follow most closely \citep[][sec.~5]{chiu22a}.

Among the model parameters of the population likelihood, there are the amplitude of the gas / stellar mass -- halo mass relation $A_{\text{ICM},\star}$, and the mass slope $ B_{\text{ICM},\star}$ of that relation. They define scaling relations that follow power laws in mass around a fixed pivot mass $M_\text{piv}$, chosen close to the median mass of the sample, reading

\begin{equation}
    \ln \Bigg(\frac{M_{\text{ICM},\star}}{M_{\text{ICM},\star}^\text{piv} }\Bigg)= A_{\text{ICM},\star} + B_{\text{ICM},\star} \ln  \Big(\frac{M_{500\text{c}}}{M^\text{piv}} \Big). 
\end{equation}
Here, we marginalize over any redshift dependencies, that are usually also studied, as they have all empirically been shown to be consistent with zero. Crucially, the statistical and systematic uncertainties incurring in the gas / stellar content and total mass measurement process, as well as in the selection of the cluster sample, are accounted for and distilled into posteriors on the scaling relation parameters $(A_{\text{ICM},\star}, B_{\text{ICM},\star})$. We therefore base our analysis on these summary statistics, together with their reported uncertainties.

\subsection{Baryon Correction Model}\label{sec:bcm}

The Baryon Correction Model proposed by \citet{schneider15} provides a semi-analytical framework to modify N-body, gravity-only simulations accounting for the effects of gas, stars and feedback. The framework is based on small, radial shifts of simulation particles around halo centres. In particular, the gravity-only profile of a given halo is mapped to the sum of a dark matter, a gas and a stellar matter profile, 
\begin{equation}
    \rho(r)\mapsto\rho_\text{DM}(r)+\rho_\text{ICM}(r)+\rho_\star(r).
\end{equation}
The model parameters determine the amplitude and shape of these profiles for each input halo mass. For each set of model parameters, the particles in the gravity-only simulation are displaced in order to obtain the new, \emph{baryonified} halo profiles. Quasi-adiabatic relaxation of the corrected profile is used to account for the gravitational back-reaction of the displaced baryons. In this work, we use two BCMs: the "S19 model" proposed by \citet{schneider15, schneider19, giri21}, and the \texttt{bacco}-model proposed by \citet{arico20, arico21a}. In spirit, the two models are similar, and we shall only quickly outline their differences in the following.

\subsubsection{S19 model}

The BCM from \citet[][hereafter S19]{schneider19} and \citet{giri21} provides a direct parametrisation of the density profiles for the gas and the stellar components. For the gas component, a power-law profile with an additional core in the centre and steep truncation in the outskirts is assumed. The functional form is given by 
\begin{equation}
    \rho_\text{ICM}(r)\propto \frac{\Omega_\text{b}/\Omega_\text{M} - f_\text{star}(M_\text{vir})}{ \left[ 1 + 10 \left(\frac{r}{r_\text{vir}} \right) \right]^{\beta(M_\text{vir})} \left[ 1 + \left(\frac{r}{\theta_\text{ej} r_\text{vir}} \right)^\gamma \right]^{ \left[\delta - \beta(M_\text{vir}) \right]/\gamma} },
\end{equation}
with a mass dependent inner slope 
\begin{equation}
    \beta(M_\text{vir}) = \frac{3\left( M_\text{vir} / M_\text{c} \right)^\mu}{1+\left( M_\text{vir} / M_\text{c} \right)^\mu},
\end{equation}
where $M_\text{vir}$ and $r_\text{vir}$ are the virial mass and the virial radius, and $\Omega_\text{b,M}$ are the cosmic baryon/matter abundance.

The stellar profile is parametrised as a central stellar component ("cga") and a satellite component ("sat"), the latter following the dark matter profile due to its collisionless nature. The fractions of the two components follow a power law in mass 
\begin{equation}
    f_\text{cga,sat} = 0.055 \left(\frac{2\times 10^{11} M_\odot /h}{M_\text{vir}} \right)^{\eta_\text{cga,sat}},
\end{equation}
with slopes $\eta_\text{sat}= \eta$ and $\eta_\text{cga}= \eta + \delta\eta$. Upon specifying the baryonic components, the collisionless profiles of the dark matter and satellite galaxies are subjected to adiabatic relaxation. This leaves us with 7 free parameters for the S19-model: $( \log_{10} M_\text{c} h / M_\odot, \mu, \theta_\text{ej}, \gamma, \delta)$ for the gas profile and $(\eta, \delta\eta)$ for the stellar component. 

Integration of the relative profiles to $r_{500\text{c}}$ allows us to compute the total mass at that over-density $M_{500\text{c}}$, as well as the enclosed gas and stellar masses $M_{\text{ICM},\star}$, from which we can readily derive the stellar and gas fractions.


\subsubsection{\texttt{bacco} model}
The \texttt{bacco} simulation projects \citep{contreras20, angulo21, zennaro23} proposed a BCM that follows a slightly different approach to define baryonic profiles. For each gravity-only halo of mass $M_{200\text{c}}$, a fraction of its mass $f_\text{ICM}(M_{200\text{c}})$ is redistributed into an empirically motivated gas distribution 
\begin{equation}
    \rho_\text{ICM}(r) \propto \begin{cases}
			\left( 1 + \frac{r}{r_\text{inn}} \right)^{-\beta_\text{inn}} \left( 1 + \left(\frac{r}{r_\text{out}}\right)^2  \right)^{-2}, & \text{ if } r< r_\text{out}\\
            \rho_\text{NFW}(r), & \text{otherwise,}
		 \end{cases}
\end{equation}
with an inner scale $r_\text{inn} = \theta_\text{inn} r_\text{200c}$, the inner slope of the gas profiles $\beta_\text{inn} = 3 - \left(\frac{M_\text{inn}}{M_\text{200c}}\right)^{\mu_\text{inn}}$, with $\mu_\text{inn}=0.31$, and the outer scale $r_\text{out} =  \theta_\text{out}^{-1} r_\text{200c}$\footnote{This definition deviates from the notation in previous works \citep{arico20, arico21a, arico21b}. It reflects the actual implementation in the code at the time of writing. Changing $\theta_\text{out}^{-1}\mapsto \theta_\text{out}$ is planned for future releases.}. $\rho_\text{NFW}(r)$ is the dark matter density profile proposed by \citet[hereafter NFW]{nfw} with the concentration mass relation fitted on the respective simulation. The normalisations of the two components are adjusted such that the transition is continuous and the integrated profiles attain the gas fraction $f_\text{ICM}(M_{200\text{c}})$ at the corresponding radius.

Another fraction $f_\star(M_{200\text{c}})$ is assumed to be stellar matter and distributed between a component mimicking the central galaxy, and another representing the satellite galaxies. This provides the prescription for the stellar matter density profile $\rho_\star(r)$. Thus, this model defines the fractions at $r_{200\text{c}}$ first, then defines the shapes of baryonic profiles.

\citet{arico20, arico21a} refined this model by allowing for a late time re-accreted gas component, and an ejected gas component. Specifically, they model the halo gas fraction within $r_{200\text{c}}$ as

\begin{equation}
    f_\text{ICM, 200c} = \Big( \frac{\Omega_\text{b}}{\Omega_\text{M}}-f_\star(M_{200\text{c}})\Big) \Bigg( 1+\Big(\frac{M_\text{c}}{M_{200\text{c}}}\Big)^\beta \Bigg)^{-1},
\end{equation}
where $M_\text{c}$ and $\beta$ are free parameters of the \texttt{bacco}-model. Note that both models define a parameter $M_\text{c}$, which however is not the same between the models. The stellar fraction is modulated by the parameter $M_1$, which is the characteristic halo mass with a stellar mass fraction of 0.023. Fitting for $M_1$ is thus equivalent to finding the halo mass, whose stellar mass fraction is 0.023. The stellar-halo mass relation is modelled with the sub-halo abundance matching proposed by \cite{Behroozi2013}, and best-fitting parameters found by \cite{Kravtsov2018}. The free parameters of the gas profile are the characteristic radii $\theta_\text{inn}$ and $\theta_\text{out}$, as well as $M_\text{inn}$, which models the inner slope of the gas profile. Finally, the parameter $\eta$ describes the position of the ejected gas cut-off. This gives a total of seven free parameters for the \texttt{bacco} model.

Contrary to the S19-model, the \texttt{bacco} model only displaces particles within a given radius $r_\text{p}$ of halos for computational ease. Deviating from previous works, which fixed $r_\text{p}=r_{\rm 200c}$, we set $r_\text{p}=r_\text{out}$. We justify this choice with the argument that beyond $r_\text{out}$ the hot gas is assumed to perfectly trace the dark matter. The only component which deviates from the NFW profile beyond $r_\text{out}$ is the ejected gas, which is going to be traced with particles displaced from inside $r_\text{out}$. We note that previous works always assumed $r_\text{out}\le r_{\rm 200c}$, thus this condition was always satisfied in previous work. This ambiguity limits the accuracy of our matter power spectrum suppression predictions to order a few percent, as discussed in more detail in App.~\ref{app:particle-radius}.

\subsubsection{Link to scaling relations}

The BCM is directly linked to the parameters of the scaling relation via a Taylor expansion of the gas / stellar mass fraction enclosed in $R_{500\text{c}}$ around the pivot mass, yielding

\begin{equation}
    f_{\text{ICM},\star}^{500\text{c}} \Big|_{M_{500\text{c}}=M_\text{piv}} = e^{ A_{\text{ICM},\star} }\frac{M_{\text{ICM},\star}^\text{piv}}{M^\text{piv}}, \text{ and}
\end{equation}

\begin{equation}
    \frac{\text{d} \ln f_{\text{ICM},\star}^{500c}}{\text{d}\ln M_{500\text{c} } }\Big|_{M_{500\text{c}}=M_\text{piv}} = B_{\text{ICM},\star} - 1 .
\end{equation}

This expression provides the link between the baryonification models and the scaling relation studies performed with Bayesian Population models. This constitutes a crucial methodological advancement with respect to the use of compilations of individual clusters that do not account for mass accuracy and selection modelling. \raul{WL calibrated Bayesian population models of cluster samples are the tool of choice to account for selection effects and mass accuracy. Compared to un-binned compilations of heterogeneously selected clusters, like the one used by \citet{flamingo2}, they are able to accurately reconstruct the underlying true halo mass distribution of the cluster samples, as well as its relation to observable quantities like the gas mass. This methodological improvement is corroborated by the fact that WL calibrated cluster population models are able to derive competitive cosmological constraints from the abundance clusters \citep{mantz15, bocquet19, chiu22cosmo}.}

Practically speaking, we also need to convert between the different mass definitions used in X-ray studies of clusters ($M_{500\text{c}}$) and the BCMs ($M_{200\text{c}}$). In the S19-model the baryon fractions at $r_{500\text{c}}$ are directly predicted by integration of the respective profiles. For the \texttt{bacco}-model the conversion from $M_{500\text{c}}$ to $M_{200\text{c}}$ is performed using the concentration mass relation by \citet{Ragagnin21} at its pivot cosmology, while the enclosed ICM mass is corrected using the gas distribution of the BCM, as further described in Appendix~\ref{app:500c-corr}.

\subsubsection{Predicting electron density profiles}\label{sec:pred_ne}

The ICM is, to good approximation, a fully ionized gas, such that the electron density profiles $n_\text{e}(r)$ can readily be estimated from the gas density as
\begin{equation}
    n_\text{e}(r) = \frac{\rho_\text{ICM}(r)}{\mu_\text{e} m_\text{p}},
\end{equation}
with the mean molecular weight of electrons $\mu_\mathrm{e}=1.17$ \citep[following ][]{bulbul19}, and the proton mass $m_\text{p}$. We can thus readily transform the ICM profiles assumed by the BCM into predictions for the electron density profiles.

One operational note is that studies of X-ray electron density profiles of massive clusters have to date not been carried out together with accurate WL mass calibration, as described for the gas and stellar fractions. We are therefore forced to rely on the hydrostatic mass reported by \citet{ghirardini19}, and transform this to a halo mass using the hydrostatic mass bias $b_\text{HS}$, with prior $p(b_\text{HS})= \mathcal{N}(b_\text{HS}| 0.26, 0.07^2)$ \citep{Hurier18}.

\subsection{Fitting Procedure}\label{sec:fitting}

\begin{table}
\caption{\label{tab:priors}
Prior choices of this analysis. All priors are chosen uniformly within the bounds that we report, except for the hydrostatic mass bias, where we use a Gaussian prior and indicate the mean and variance of the distribution. Note that both BCMs use the parameter $\log_{10} M_\text{c}h / M_\odot$, but it has different meanings in the two models, as described in section~\ref{sec:bcm}. }
\begin{tabular}{lclc}
\multicolumn{2}{c}{AS-model} & \multicolumn{2}{c}{\texttt{bacco}} \\
\hline
$\log_{10} M_\text{c}h / M_\odot$ & $(11,\,15)$ & $\log_{10} M_\text{c}h / M_\odot$ & $(12,\,15)$ \\ 
$\mu$ & $(0,\,2)$ & $\log_{10} \beta$ & $(-1,\,0.7)$ \\
$\theta_\text{ej} $ & $(2,\,8)$ & $\log_{10} M_1 h / M_\odot$ & $(11.5,\,15)$ \\
$\gamma$ & $(1,\,4)$ & $\log_{10} \eta$ & $(-0.7,\,0.7)$ \\
$\delta$ & $(3,\,11)$  & $\log_{10} \theta_\text{out}$ & $(-1,\,0.5)$ \\
$\eta$ & $(0.05,\,0.4)$ & $\log_{10} M_\text{inn}h / M_\odot  $ & $(5,\,15)$ \\
$\delta\eta$ & $(0.05,\,0.4)$ & $\log_{10} \theta_\text{inn}$ & $(-2,\,0)$ \\
\hline
\multicolumn{4}{c}{cosmological parameters} \\
$\Omega_\text{M}=0.315$ & $h=0.67$ & \multicolumn{2}{c}{$\Omega_\text{b}\in (0.0473, 0.0535)$} \\
\hline
\multicolumn{4}{c}{hydro static mass bias} \\
\multicolumn{4}{c}{$b_\text{HS} \sim \mathcal{N}(0.26, 0.07^2)$} \\
\end{tabular}
\end{table}

\begin{figure*}
	\includegraphics[width=\columnwidth]{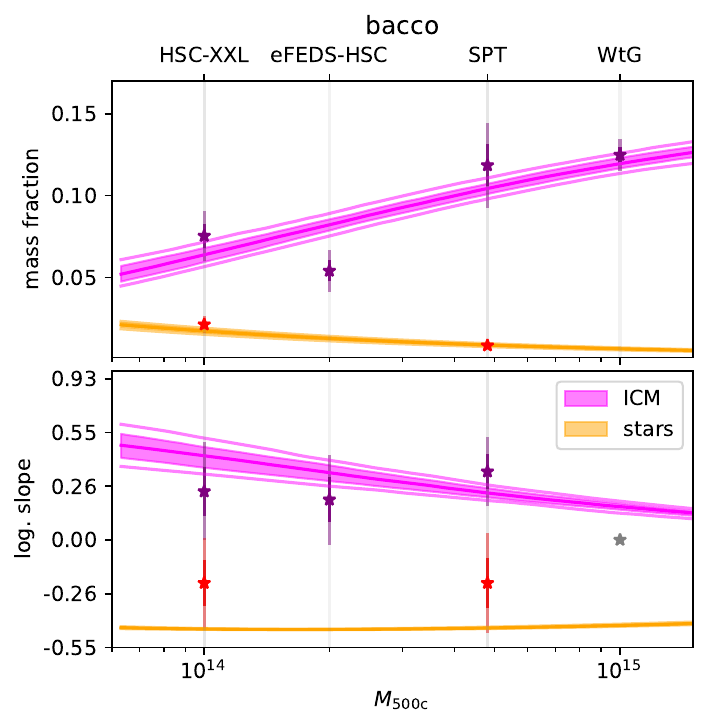}
    \includegraphics[width=\columnwidth]{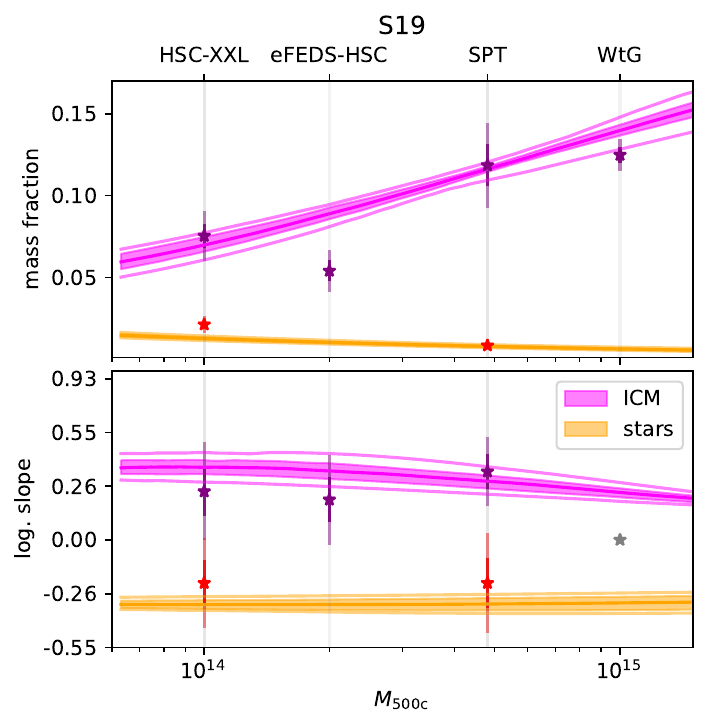}
    \caption{$\textit{Top:}$ Prediction for the gas (magenta) and stellar (orange) mass fraction together with the fitted data points from the cluster analyses considered. We show the 1 sigma region as filled and represent also the 2 sigma region with faded lines. $\textit{Bottom:}$ Prediction for the mass slope of the gas (magenta) and stellar (orange) mass fraction together with the fitted data points from the cluster analyses considered. \rvs{The data points for the slope are measured by cluster population studies and are statistically independent of the baryon fraction measurements. The model for the slope is the log-derivative w.r.t. to log-mass of model predictions for the fractions.} In grey the slope measurement we discard due to its implausibly tight error bar. The left panel shows the \texttt{bacco} fit, while the right shows the S19-model fit. \rvs{Due to anomalies in the data, the BCMs are only able to capture the qualitative trends in the data}.} 
    \label{fig:fit}
\end{figure*}

We sample a Gaussian likelihood for the measurements reported in  Table~\ref{tab:fracs} and Table~\ref{tab:ne_prof}. The mean values of the individual data point are given by the model prediction of the BCMs, and the respective variances are given by the square of the measurement uncertainties reported above (cf. section~\ref{sec:data}, Table~\ref{tab:fracs}, and Table~\ref{tab:ne_prof}). The measurements are treated as independent, as justified in section~\ref{sec:data}. The model prediction for the gas and stellar fraction from the S19-model is computed by integrating the respective profiles to $R_\text{500c}$, while the prediction of the $\texttt{bacco}$ model relies on a combination of functions provided by the package $\texttt{bacco}$\footnote{\url{https://bacco.dipc.org/}} \citep{arico21b} and the necessary conversions described in section~\ref{sec:bcm}. Predictions for the electron density profiles directly result from appropriate re-scaling of the gas density profile (cf. Section.~\ref{sec:pred_ne}).

We employ the Monte Carlo Markov Chain sampler $\texttt{emcee}$\footnote{\url{https://emcee.readthedocs.io/en/stable/}} \citep{emcee} to sample the posteriors. We sample all 7 parameters of the BCMs within flat priors (for the exact ranges, see Table~\ref{tab:priors}). The cosmic baryon density $\Omega_\text{b}$ is sampled in all cases within the prior ranges $(0.0473, 0.0535)$ to emulate a residual uncertainty in the cosmic baryon fraction, which impacts the gas mass fraction predictions and the amplitude of the predicted electron density profiles.

The cosmological parameters are kept fixed at $\Omega_\text{M}=0.315$, and $h=0.67$. The choice of cosmic baryon density $\Omega_\text{b}$ reflects the range of cosmic baryon fractions sampled by \citet{schnieder21}, $0.15<f_\text{b}<0.17$. The Hubble constant is kept fixed as our analysis is, to first order, independent of the Hubble parameter. The BCMs require input masses in units of $M_\odot$/$h$ by design. Observations only have access to the angular scale of clusters or groups. As such, even if not explicitly reported, observational masses are effectively in units of $M_\odot$/$h$. Similarly, the electron density profile is effectively measured $h$-scaled units, matching the density profile predictions. To first order, this analysis thus depends on the main cosmological parameters only via the cosmic baryon fraction. In future work, especially when used as priors for cosmological analyses, we plan to take account also of the residual cosmological dependence.

\section{Results}

In this section, we first discuss the posteriors of our analysis, both with regard to the model parameters themselves, as well as the posterior predictive distributions of the data. We then derive the posterior predictive distributions for the matter power spectrum suppression due to baryonic effects.

\subsection{Parameter Posteriors and Goodness of Fit}\label{sec:posterior&gof}

\begin{figure}
	\includegraphics[width=\columnwidth]{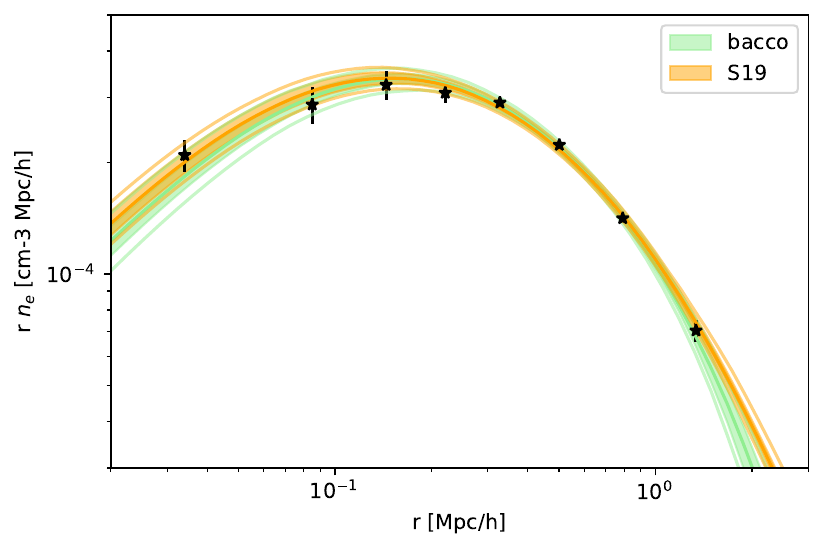}
    \caption{Fit to the electron density profile (black points), re-scaled by the radius to reduce the dynamical range of the plot. Color coded the different BCM fits. \rvs{Computing the chi-squared of the median posterior predictions and the data, we find that b}oth models provide a good fit to the data.}
    \label{fig:ne_fitted}
\end{figure}

\begin{table}
\caption{\label{tab:priors}
Mean and standard deviations of the parameter posteriors if the parameter is well measured. For parameters with lower (upper) limits we report the 2.5\% (97.5\%) percentiles. Note that both BCMs use the parameter $\log_{10} M_\text{c}h / M_\odot$, but it has different meanings in the two models, as described in section~\ref{sec:bcm}. }
\begin{tabular}{lclc}
\multicolumn{2}{c}{AS-model} & \multicolumn{2}{c}{\texttt{bacco}} \\
\hline
$\log_{10} M_\text{c}h / M_\odot$ & $14.53 \pm 0.20$ & $\log_{10} M_\text{c}h / M_\odot$ & $13.82 \pm 0.36$ \\ 
$\mu$ & $0.54\pm 0.10$ & $\log_{10} \beta$ & $-0.29 \pm 0.11$ \\
$\theta_\text{ej} $ & $< 6.72$ & $\log_{10} M_1 h / M_\odot$ & $12.00\pm 0.06$ \\
$\gamma$ & $<1.42$ & $\log_{10} \eta$ & -- \\
$\delta$ & $>5.37$  & $\log_{10} \theta_\text{out}$ & $< -0.11$ \\
$\eta \left( \delta\eta / 0.24\right)^{0.19}$ & $0.239\pm 0.009$ & $\log_{10} M_\text{inn}h / M_\odot  $ & -- \\
$\delta\eta$ & -- & $\log_{10} \theta_\text{inn}$ & $ < -0.65$ \\
\hline
\end{tabular}
\end{table}

When constructing a posterior sample, we discard the burn-in phase of our sampling and visually inspect the trace plots to ensure convergence, especially the trace plot of the total log-likelihood. For sake of brevity, the 1- and 2-d marginal posteriors are reported in Appendix, Fig.~\ref{fig:full_post_bacco} and Fig.~\ref{fig:full_post_AS}. 

For the \texttt{bacco} model, we find $\log_{10} M_\text{c}h / M_\odot = 13.82 \pm 0.36$, $\log_{10} \beta = -0.29 \pm 0.11$, and $\log_{10} M_1 h / M_\odot=12.00\pm 0.06$. For the parameters defining the scales of the gas profile, we only find upper limits:  $\log_{10} \theta_\text{inn} < -0.65$ and $\log_{10} \theta_\text{out} < -0.11$ (at the 97.5th percentile). The position of the expelled gas $\eta$, and the characteristic mass of the inner gas profile slope $M_\text{inn}$ remain unconstrained. As seen in the 1- and 2-d marginal contours shown in Fig.~\ref{fig:full_post_bacco}, we find a major degeneracy between the characteristic mass of the inner gas profile slope $M_\text{inn}$ and corresponding scale $\log_{10} \theta_\text{inn}$.

For the S19 model, we find $\log_{10} M_\text{c}h / M_\odot =14.53 \pm 0.20$, and $\mu = 0.54\pm 0.10$. We get upper limits on $\theta_\text{ej}< 6.72$, and $\gamma<1.42$ (at the 2.5th percentile); and a lower limit on $\delta>5.37$ (at the 97.5th percentile). The mass slopes of the satellite stellar mass $\eta$, and the difference between the satellite and central stellar mass slopes $\delta\eta$ are strongly degenerate with each other (see Fig.~\ref{fig:full_post_AS}). The best constrained parameter combination is $\eta \left( \delta\eta / 0.24\right)^{0.19}=0.239\pm 0.009$.

In Fig.~\ref{fig:fit} we present the model prediction for the ICM (magenta) and stellar (orange) mass fraction (upper panels) as a function of mass over plotting the data points used for the fitting. Similarly, in the lower panel of Fig.~\ref{fig:fit} we show the model prediction for the logarithmic mass slope of the ICM and stellar mass fraction (same color scheme as upper panel), together with the slope measurements. The left images show the best fit \texttt{bacco} model for the baryon fractions and their mass slopes. The right plot shows the baryon fraction and mass slope predictions for the S19-model. Both models that we investigate are able to capture the main trends, \rvs{while the overall quality of the fit is hampered by anomalies in the data.}

A closer inspection reveals a few anomalies. In grey we plot the discarded slope measurement by \citet{mantz16}, which has implausibly small error bars, as can be seen in relation to the other analyses. Furthermore, the eFEDS-HSC ICM fraction measurement seems to be several $\sigma$ low with respect to the predicted range of gas fractions at that mass, independently on the BCM. We discuss this below (cf. Section~\ref{sec:disc}). The model predicts very accurately the logarithmic slope of the ICM mass fraction. The logarithmic mass slope of the stellar mass fraction from the \texttt{bacco} model is almost 2 sigma lower than our measurements. Conversely, the AS model seems to better predict the stellar mass fraction slopes, but at the cost of under predicting the stellar fraction of the HSC-XXL analysis. We discuss these minor anomalies below (cf. Section~\ref{sec:disc}).

\rvs{The fit of both models fares better on the electron density data, shown in Fig.~\ref{fig:ne_fitted}. Here the median posterior predictive derived from the \texttt{bacco} model attains a chi-squared of $\chi^{2}=3.8$ on the 8 data points of the electron density profile, while effectively constraining 3 model parameters and having a tight degeneracy on another pair of parameters. For the S19 model, the median posterior predictive has a chi-squared of $\chi^{2}=4.8$ on the 8 data points of the electron density profile, while effectively constraining 2 model parameters, placing two limits on other parameters, and having a tight degeneracy on another pair of parameters. The chi-squared values of the median model predictions are in line with the expectation from the effective number of degrees of freedom. This means that the BCM are able to fit the electron density data well. }

\subsection{Impact on the matter power spectrum}

\begin{figure*}
	\includegraphics[width=\textwidth]{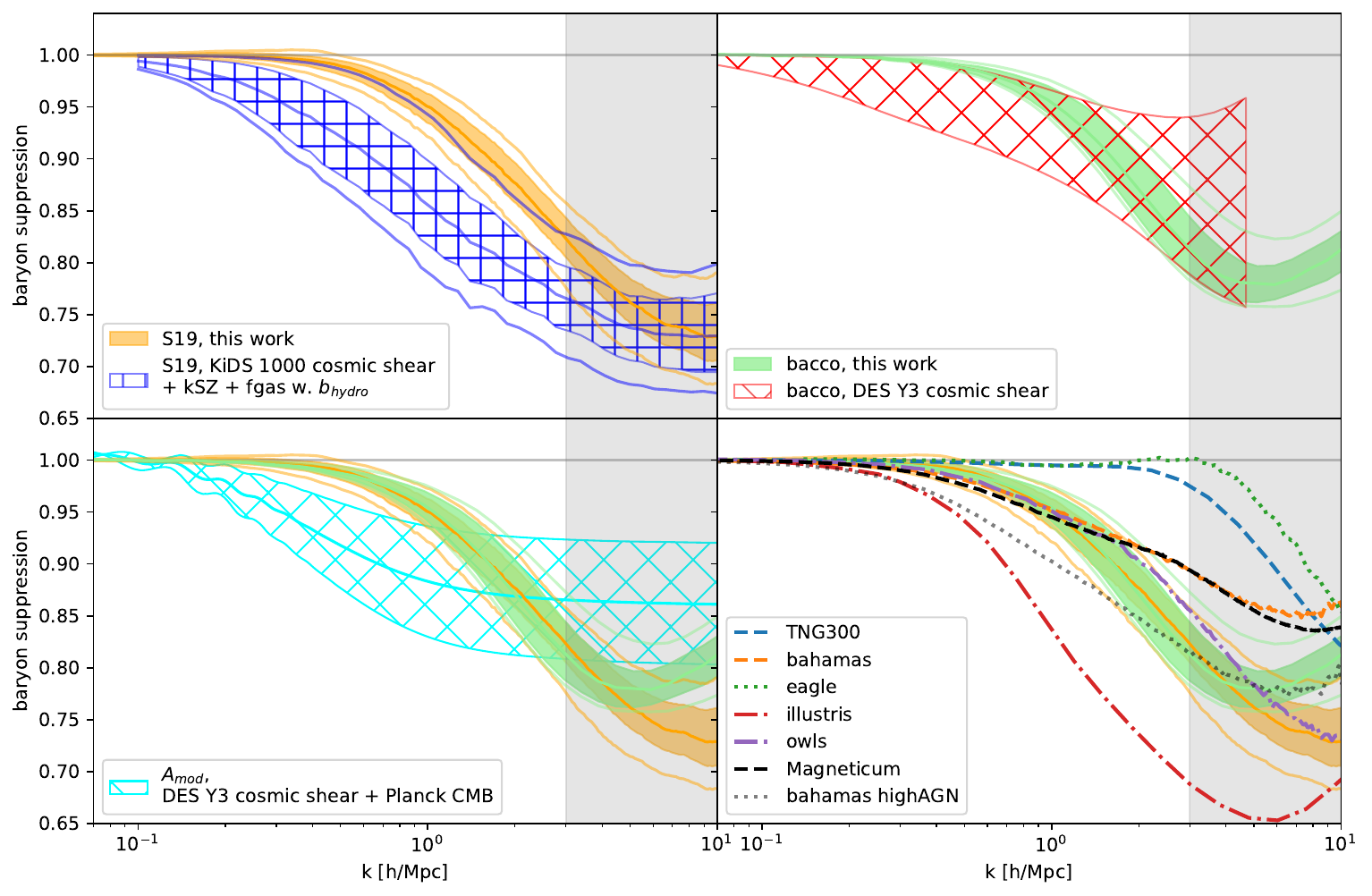}
    \caption{Posterior predictive distribution of the suppression in the matter power spectrum, that is the ratio of the matter power spectrum with baryonic effects, and the gravity only matter power spectrum, $P_{\rm with \, baryons}/P_{\rm gravity \, only}$, obtained with the S19 and \texttt{bacco} models (orange and green shaded area, respectively) informed with the galaxy cluster data collected in this work. The full lines represent the 2.5th, 50th, and 97.5th percentile at each wave-number $k$, while the filled regions encompass the range between the 16th and 84th percentile. Greyed out the scales impacted by extrapolating our data to lower halo masses.
    \textit{Upper left:} Comparison with S19-model prediction from KiDS 1000 cosmic shear, kSZ measurements and individual cluster with hydrostatic masses \citep[hatched blue]{schnieder21}, which find a two sigma stronger suppression. \textit{Upper right:} Comparison with the \texttt{bacco} prediction from DES Y3 cosmic shear \citep[hatched red]{arico23}, which is in good statistical agreement. \textit{Lower left:} in cyan the suppression needed to reconcile DES Y3 cosmic shear and Planck primary CMB constraints \citep[a.k.a. S8 tension, hatched in cyan,][]{preston+23}. Predictions based on galaxy clusters can not explain the large scale suppression needed to solve the S8 tension. \textit{Lower right:} the prediction of several state-of-the-art hydrodynamical simulations, according to the legend. Our data suggest an intermediate to strong feedback scenario.
    }

    \label{fig:hydro_suppr}
\end{figure*}

Given our posterior of the BCM parameters, we can predict the suppression of the matter power spectrum induced by the evacuation of baryons from massive halos. Note that our upper limit on the outer characteristic scale of the \texttt{bacco} model ($\theta_\text{out}<-0.11$) falls outside of the range of the $\texttt{bacco}$ emulator \citep[$0<\theta_\text{out}<0.5$,][]{arico21b}. We therefore recompute the matter power spectrum suppression for this model for $\mathcal{O}(1000)$ points drawn from our posterior. For the S19-model, we use the \texttt{BCemu} \citep{giri21} to compute the baryon suppression of the matter power spectrum. The resulting posterior predictive distributions are shown in Fig.~\ref{fig:hydro_suppr}, with the full line representing the 2.5th, 50th, and 97.5th percentile at each wave-number $k$, while the filled regions encompass the range between the 16th and 84th percentile. In light green we show the suppression for the \texttt{bacco}-model, while the orange shows the suppression for the S19-model. Via the BCMs, our data is able to predict matter power spectrum suppressions on all scales of interest with noteworthy precision.

We find remarkable agreement  between the matter power suppression predictions from the two models on scales larger than $k< 6h$ Mpc$^{-1}$, considering that the two models are completely independent implementations with different parametrizations. Both models quickly converge towards no suppression at scales larger than  $k< 0.5h$ Mpc$^{-1}$. Specifically, for $k<0.37h$ Mpc$^{-1}$ ($k<0.29h$ Mpc$^{-1}$) 97.5 percent of our predictive posterior shows as suppression weaker than 1\% for the \texttt{bacco} (S19) model, indicating that baryon feedback processes do not impact the matter power spectrum at larger scales. 

The two BCMs also show the same strong suppression trend at intermediate scales, $0.5h$ Mpc$^{-1}$ $<k < 6 h$ Mpc$^{-1}$. The predicted suppression is still rather weak at $k=1h$ Mpc$^{-1}$: $0.042^{+0.012}_{-0.014}$ for \texttt{bacco}, and $0.049^{+0.016}_{-0.012}$ for the S19-model. The suppression grows sizeably at $k=3 h$ Mpc$^{-1}$, predicting $0.184^{+0.026}_{-0.031}$ for \texttt{bacco}, and $0.179^{+0.018}_{-0.020}$ for the S19-model.  At smaller scales, the \texttt{bacco} prediction has the characteristic up-turn at larger scales than the S19 prediction. The prediction of the two models thus diverges. At $k=10 h$ Mpc$^{-1}$ we find $0.185^{+0.023}_{-0.017}$ for \texttt{bacco}, and $0.271^{+0.023}_{-0.033}$ for the S19-model. 
The discrepancy at small scales is to be expected given our lack of data for $M_\text{500c} < 10^{14} M_{\odot}$.  \citet{arico21a} highlight that halos in the mass range $10^{14} h^{-1} M_\odot < M_\text{200c} < 10^{15} h^{-1} M_\odot$, contribute most to the matter power spectrum suppression above scales of $k\lesssim 3 h$ Mpc$^{-1}$, while halos in the mass range of $10^{13} h^{-1} M_\odot < M_\text{200c} < 10^{14} h^{-1} M_\odot$ contribute the most to scales  $k\gtrsim 3 h$ Mpc$^{-1}$. We highlight this by shading the area in Fig.~\ref{fig:hydro_suppr} that we do not directly constrain with our data. Discrepancies in the stellar mass fraction results might also play a role in the divergence of the two model predictions, as discussed below. Also, the accuracy of these predictions at large scales is impacted by the details of the particle displacement procedure in the BCMs at percent level (see App.~\ref{app:particle-radius}). More in-depth work is required to understand the source of these differences and improve further the agreement among the various baryonification methods.

\section{Discussion}\label{sec:disc}

In the following, we will discuss several aspects of this work.

\paragraph*{Weak lensing masses} This work provides a clear improvement compared to prior attempts to use the gas mass -- halo mass relation of galaxy clusters and groups to constrain the baryonic suppression of the matter power spectrum \citep{schneider19, debackere20, giri21}, which use hydrostatic mass estimates to interpret baryon fractions. Those works showed that the limiting factor in using hydrostatic masses was the weakly constrained hydrostatic mass bias. Also recent empirical calibrations of hydro-dynamical simulations are limited by the use of hydrostatic masses \citep{flamingo2}. This uncertainty is reduced by using weak lensing mass calibration for the baryon fractions. Furthermore, the cluster and group measurements in those works did not account for selection effects via Bayesian population models, while the analyses used in this work do. In future work, we plan to also use weak lensing mass calibration and model selection effects also for the exploitation of electron density profiles.

\paragraph*{Comparison to previous results} \rvs{Previous work by \citet{giri21, schnieder21} fitted the S19 model with various data sets. While neither of the two works resulted in a constraint on any of the BCM parameters, their posteriors suggest that $\log_{10} M_\text{c}<14$, which is in slight tension with our measurement. Both works also indicate that $\mu \sim 0.4-0.5$, though at low significance. Our measurement $\mu = 0.54\pm 0.10$ is in agreement with that. Also our constraint that $\eta = 0.239\pm 0.009$ at $\delta\eta = 0.24$ coincides with the posterior distributions they plot.} \citet{arico23} recently analysed the Dark Energy survey year 3 cosmic shear auto-correlation with the $\texttt{bacco}$ model. They find a value of $\log_{10} M_\text{c} h / M_\odot=14.38^{+0.60}_{-0.56}$ -- in  agreement with our measurement. We therefore confirm the best fit small scale baryonic effects found by \citet{arico23} when analyzing cluster gas density profiles, and gas and stellar fractions. Our results are also in excellent agreement with the analysis by \citet{chen+22} on cosmic shear data only, \rvs{though in that work, 6 of the 7 parameters were fixed to the best fit values from the BAHAMAS simulation}. Compared to those works, which only constrain one BCM parameter, we are able to measure 3 out of 7 parameters tightly, while placing upper limits on the other 2. In Fig.~\ref{fig:hydro_suppr}, upper panels, we directly compare our matter power spectrum suppression predictions with the ones from the \texttt{bacco} fit to DES Y3 cosmic shear \citet{arico23} (bacco, DES Y3 WL in hatched red), and from the fit of the S19-model to KiDS cosmic shear, kSZ measurements and a compilation of gas and stellar fractions \citep{schnieder21}. While the former is in excellent agreement with our results, the prediction by \citet{schnieder21} is 2-sigma lower on all scales. If this is due to our inclusion of the X-ray profiles, or their use of kSZ data, is left to future investigation.

\paragraph*{S$_8$ tension} Current cosmological constraints from wide photometric surveys suggest that the amplitude of fluctuation in the low redshift Universe is lower than predicted by extrapolating cosmic Microwave background experiments \citep[see][for a review]{huterer23}.
In the context of cosmic shear only, it reaches a significance of $1.6\sigma$ in the joint Dark Energy Survey + Kilo Degree Survey analysis \citep{des+kids}, while the significance is $2\sigma$ in the latest Hyper Supreme Camera work \citep{Li2023,Dalal2023}. 
The recent CMB lensing analysis of ACT \citep{act-cmbl}, which is in agreement with primary CMB predictions, \rvs{indicates that this tension might not be present at large scales and redshift larger than  $\sim 1$. As pointed out by \citet{amon22, preston+23}, the solution would likely have to be non-linear, and at low redshift. To elaborate on this, we} compare the matter power spectrum suppression derived by redistributing baryons in and around halos, with the suppression needed to reconcile DES Y3 cosmic shear and Planck primary CMB constraints, derived by \citet{preston+23}, whose 1-sigma region is shown in cyan in Fig.~\ref{fig:hydro_suppr}, lower left panel. That work, as well as a comparable analysis with KiDS cosmic shear \citep{amon22}, find that a solution to the "S8 tension" would require strong suppression of large scale, $0.2 h$ Mpc$^{-1}<k<1 h$ Mpc$^{-1}$. Given our predictions, it seems unlikely the baryon feedback can provide such suppression by redistributing baryons in and around halos, even when considering the modelling uncertainties discussed in App.~\ref{app:particle-radius}. Using the approximation $k\approx \pi/R$, solving the S8 tension would require a suppression at scales of from $3 h^{-1}$ Mpc to $15 h^{-1}$ Mpc, far outside even the outermost regions of halos. Following the interpretation by \citet{amon22}, this would hint at yet unknown physical effects impacting the dark matter at (mildly) non-linear scales. This also confirms that in the context of cosmic shear, appropriate scale cuts can make experiments impervious to baryonic effects by only considering large scales \citep{krause21, arico23, des+kids}. Previous constraints on BCM confirm that baryon feedback only plays a subordinate role in solving the S8 tension. Indeed, a re-analysis of the Kilo Degree Survey 1000 cosmic shear data by \citet{schnieder21} using the S19 model, shows that the baryon feedback can not resolve, and only marginally alleviate the discrepancy between cosmic shear and CMB.
A re-analysis of DES Y3 data by \citet{arico23} find similarly that baryons have a marginal effect on the $S_8$-tension when using scale-cuts. However, when combined with other effects e.g. non-linear and intrinsic-alignment models, they can have an impact of more than 1.4$\sigma$, pointing out that the tension might also arise by shortcomings in the modelling.

\paragraph*{Hydro-dynamical simulations} We compare in Fig.~\ref{fig:hydro_suppr}, lower right panel, our predicted matter power suppression to several point estimates from state-of-the-art hydro-dynamical simulations: Magneticum \citep[][and references therein]{martinet21}, TNG300 \citep{pillepich18,  springel18}, BAHAMAS \citep{maccarthy17, mccarthy18}, EAGLE \citep{Schaye2015,Crain2015,McAlpine2016}, Illustris \citep{vogelsberger10}, and OWLS \citep{schaye10}.  Both very strong feedback models (like the Illustris simulation), and very weak feedback models (like TNG300 and EAGLE) are effectively outside of the combined range of our constraints, even when considering the variation introduced by the difference between the baryonification models. Our posteriors qualitatively follow \rvs{most closely} the trend of the OWLS simulations. Our work further supports the strong feedback scenario previously empirically suggested by \citet{eckert16, giri21}.

\paragraph*{Selection effects} The poor fit the BCM provides to the gas fraction of eFEDS-HSC (see Fig.~\ref{fig:fit}, upper panel) could be indicative of some of the challenges faced by Bayesian Population studies of galaxy clusters and groups. Objects of the same mass with higher gas mass fraction would plausibly also be more X-ray luminous than objects that have retained, at the same mass, less baryons \citep[for simulational evidence, see][]{ragagnin22}. \citet{popesso23} have shown, that X-ray selected groups are more concentrated, both in X-rays as well as in total mass. They also preferentially live in nodes of the cosmic web and have redder central galaxies than groups detected only via friends-of-friends algorithms in spectroscopic galaxy data. In a Bayesian Population model, this can be accounted for by a correlation between the intrinsic scatters of different observables \citep{angulo12, mantz15, bocquet15, farahi19, grandis21}. While \citet{akino22} did not explicitly fit for a possible correlation between selection observable scatter and gas mass scatter, \citet{chiu22a} find that such correlation is consistent with zero, albeit with large error bars. 
\rvs{Both these analyses will be soon superseded by the publication of the first eROSITA all sky survey \citep{predehl21}. In that context, multi wavelength cross checks to validate the selection modelling, such as proposed by \citet{grandis20, grandis21}, should be performed. We plan to revisit the predictions of the matter power spectrum suppression once low mass halo selection is better understood.}

\paragraph*{Internal consistency} Further anomalies can be found in our compilation of data sets. For instance, if one was to take the two measurements of the stellar mass fraction as nodes for a linear interpolation, one would find, after accounting for errors, a stellar fraction slope of $B_\star-1 = -0.58 \pm 0.10$. This is more than 2$\sigma$ (and less than 3$\sigma$) steeper than the measured slopes $B_\star-1 = -0.20 \pm 0.11$, and $B_\star-1 = -0.20 \pm 0.12$, from HSC-XXL, and SPT respectively. This explains why neither of our models is able to simultaneously fit the stellar mass fractions and their mass slopes. Such inconsistencies could be avoided by fitting larger future cluster and group samples directly with the prescriptions from baryon correction models. The relative stiffness of the BCMs also results in rather low stellar fractions, especially in the S19 model. This is likely the reason for the different behaviours of the BCM predictions at small scales. Less stellar material would imply more baryons as hot gas. In light of the measured gas fraction, it would seem that we do not find this excess hot gas in the halos. To reconcile this, we would need strong feedback. In summary, in our inference set-up, under-predicted stellar fractions lead to more feedback.

\paragraph*{Mass measurement accuracy} Another limiting factor for the accuracy of gas and stellar mass fraction measurements for galaxy clusters and groups is the accuracy of the weak lensing mass estimation. Many of the observational systematics in shape and photometric redshift measurement, as well as background selection will be alleviated by the increased data quality of future data provided by surveys like \textit{Euclid} and LSST. \citet{grandis21b} however showed that also the theoretical knowledge of the total matter distribution of halos can lead to systematic uncertainties no smaller than 2\% \citep[see also][]{debackere21}. This systematics floor results from the comparison between Magneticum and TNG300. Given that the weak feedback scenario in the latter is excluded by this work, the actual uncertainty on baryon feedback might well be smaller than the reported 2$\%$. In this context data constrained BCMs, like the one we present here, can be used to marginalise over the residual theoretical uncertainty in baryon feedback effects on the matter profiles of galaxy clusters.

\paragraph*{Modelling improvements} At a qualitative level the performance of the BCMs in fitting our data satisfies us, so we see no immediate source of concern. It is also noticeable that the high precision of the used data can be effectively propagated by the BCMs to high precision matter power spectrum suppression predictions. Concerning the predictive accuracy of the BCMs, \citet{schneider19, giri21, schnieder21} explicitly show that when fitting the baryon fraction measured in different hydro-dynamical simulations, the S19-model is able to correctly predict the matter power spectrum suppression, and vice versa. While the S19 model is also able to fit gas density profiles derived from X-ray observations, it has not been shown explicitly that fits to the electron density profiles and baryonic fraction correctly predict the matter power spectrum. For the \texttt{bacco}-model, \citet{arico21a} have shown that it is able to jointly fit the matter power spectrum and bispectrum in hydrodynamical simulations, and thereby reproduce baryon fractions. Separate fits are able to describe the gas and stellar density profiles.  The predictive accuracy on the matter power spectrum of fits to baryon fraction and electron density profiles remains to be tested. We highlight in App.~\ref{app:particle-radius} how the details of the particle displacement prescription impact the matter power spectrum suppression at percent level. We will further investigate in future works the accuracy of the baryonification as a function of the displacement radius. Similarly, we will also study the redshift evolution of baryon feedback, both in the collection of data, as well as the modelling. 

\paragraph*{Other data sets} While cluster baryon fractions are clearly a potent way to empirically constrain baryon feedback, they only put constraints on a rather narrow range of scales, given the limited mass range they probe. For scales $k\gtrapprox1 h$ Mpc$^{-1}$, the X-ray surface brightness profile of clusters provide further empirical constraint, as shown in this work. We only used a small fraction of the recorded data. It will surely be fruitful to ingest more X-ray data, especially at lower halo masses, into the BCM fitting procedure, an effort we hope this work provides the motivation for.  At even smaller scales, the relation between central galaxy stellar mass and host halo mass for low mass groups will be crucial. Also here, ample observational information exists, though, in the authors' opinion, rarely in an easy to access format. For both of these observational channels, the challenge is to interface existing and ongoing observations with the BCM. It remains to be tested if data covering a wider mass range would be well fitted by the BCM, or if the current extrapolation to lower masses leads to artificially precise, but inaccurate prediction.
On scales larger than $k<2 h$ Mpc$^{-1}$, the hot gas can be traced with the Sunyeav-Zeldovich effect (SZe), as recently shown by \citet{pandey19, desi-groups, gatti22, troester22} for thermal SZe. \citet{schnieder21, chaves-montero21} showed constraints from kinematic SZe, which requires fewer modelling assumptions, but is also lower signal to noise than thermal SZe. The dispersion of localized fast radio bursts also probes the cosmic baryon fraction \citep{macquart20}, and future larger samples of localised fast radio bursts will be able to probe the baryon distribution \citep{nicola22}. Another approach is to directly exploit the high signal to noise of small scale galaxy and cosmic shear lensing to constrain the matter power spectrum suppression \citep{yoon21, huang21, amon22,chen+22,  preston+23, arico23}, with the disadvantage that it would not be clear, if such a suppression signal is physically due to baryon feedback, or some other physically interesting effect on non-linear scales. The tightest and most robust constraints will naturally result from the combination of all these observations.

\section{Conclusions}

In this work, we use gas and stellar mass fraction measurements derived by several weak lensing informed multi wavelength galaxy cluster studies \citep{mantz16, chiu18, chiu22a, akino22}, as well as the electron density profile derived by deep, high resolution X-ray observations \citep{ghirardini19}, to empirically inform the Baryon Correction Model proposed by \citet{arico20, arico21a} and by \citet{schneider15, schneider19}. \rvs{Despite some anomalies in the observed data,} we find that the data points used are \rvs{qualitatively} described by the proposed models, once the free parameters of these models are constrained by Bayesian inference. Using our constraints on the BCM parameters, we predict the suppression of the non-linear matter power spectrum with sub-percent precision. We find that at scales smaller than $k<0.37h$ Mpc$^{-1}$ the baryon suppression is $<1\%$ at 97.5 percent credibility. At $k=1 h$ Mpc$^{-1}$ we find a suppression of $0.042^{+0.012}_{-0.014}$ for \texttt{bacco}, and $0.049^{+0.016}_{-0.012}$ for the S19-model, while at $k=3 h$ Mpc$^{-1}$, we find $0.184^{+0.026}_{-0.031}$ for \texttt{bacco}, and $0.179^{+0.018}_{-0.020}$ for the S19-model. The predictions diverge at small scales for $k>5h$ Mpc$^{-1}$ most likely due to data anomalies driving different model fits, and our limited range in halo mass leading to extrapolation inaccuracies. 

Given our predictions, we can exclude a series of simulations that feature both too strong, or too weak feedback effects. Our predicted matter power suppression seems to be reproduced most closely by the hydro-dynamical simulation results from OWLS \citep{schaye10}, \rvs{while other simulations deviated from our posterior predictive on small or large scales}.  Constraints on the baryonic effects derived from small scale cosmic shear by \citet{chen+22, arico23} are empirically confirmed by our fit to stellar and gas mass fractions, as well as electron density profiles. Comparing our results with the matter power spectrum suppression needed to solve the S8 tension \citep{amon22, preston+23}, we find that baryon feedback can likely not provide the necessary large scale suppression, and therefore is unlikely to solve the S8 tension.

In summary, we demonstrate that current measurements of galaxy cluster gas and stellar fraction, and cluster gas profiles can  strongly constrain the baryon correction to the matter power spectrum at percent level. The cluster data can thus act as ancillary data products, calibrating astrophysical uncertainties in cosmic shear experiments. The current predictive precision is found to be of the same order of magnitude as the predictive accuracy of the BCM, suggesting that future work should investigate the differences that arise among the various BCMs / displacement algorithms, and test which one provides the most accurate predictions. This would open up the possibility of using small scale cosmic shear measurement to constrain deviation from the standard cosmological model, while controlling the baryonic effects via external data, if other systematics are similar well constrained. The improved understanding of baryon feedback will also reduce theoretical uncertainties in weak lensing calibrated cluster number counts.

\section*{Acknowledgements}

The authors thank Sebastian Bocquet, Inon Chiu, Tim Schrabback, Daisuke Nagai, Vittorio Ghirardini, Dominique Eckert, Alexandra Amon and Raul Angulo, as well as the eROSITA Cluster Working Group for the useful comments provided at different stages of this work. Marginal contour plots of high-dimensional samples are visualized with \texttt{pyGTC} \citep{Bocquet2016}.

\section*{Data Availability}

All data will be made available upon reasonable request to the authors.



\bibliographystyle{mnras}
\bibliography{example} 

\begin{thebibliography}{}
\makeatletter
\relax
\def\mn@urlcharsother{\let\do\@makeother \do\$\do\&\do\#\do\^\do\_\do\%\do\~}
\def\mn@doi{\begingroup\mn@urlcharsother \@ifnextchar [ {\mn@doi@} {\mn@doi@[]}}
\def\mn@doi@[#1]#2{\def\@tempa{#1}\ifx\@tempa\@empty \href {http://dx.doi.org/#2} {doi:#2}\else \href {http://dx.doi.org/#2} {#1}\fi \endgroup}
\def\mn@eprint#1#2{\mn@eprint@#1:#2::\@nil}
\def\mn@eprint@arXiv#1{\href {http://arxiv.org/abs/#1} {{\tt arXiv:#1}}}
\def\mn@eprint@dblp#1{\href {http://dblp.uni-trier.de/rec/bibtex/#1.xml} {dblp:#1}}
\def\mn@eprint@#1:#2:#3:#4\@nil{\def\@tempa {#1}\def\@tempb {#2}\def\@tempc {#3}\ifx \@tempc \@empty \let \@tempc \@tempb \let \@tempb \@tempa \fi \ifx \@tempb \@empty \def\@tempb {arXiv}\fi \@ifundefined {mn@eprint@\@tempb}{\@tempb:\@tempc}{\expandafter \expandafter \csname mn@eprint@\@tempb\endcsname \expandafter{\@tempc}}}

\bibitem[\protect\citeauthoryear{{Akino} et~al.,}{{Akino} et~al.}{2022}]{akino22}
{Akino} D.,  et~al., 2022, \mn@doi [\pasj] {10.1093/pasj/psab115}, \href {https://ui.adsabs.harvard.edu/abs/2022PASJ...74..175A} {74, 175}

\bibitem[\protect\citeauthoryear{{Amon} \& {Efstathiou}}{{Amon} \& {Efstathiou}}{2022}]{amon22}
{Amon} A.,  {Efstathiou} G.,  2022, \mn@doi [\mnras] {10.1093/mnras/stac2429}, \href {https://ui.adsabs.harvard.edu/abs/2022MNRAS.516.5355A} {516, 5355}

\bibitem[\protect\citeauthoryear{{Angulo}, {Springel}, {White}, {Jenkins}, {Baugh}  \& {Frenk}}{{Angulo} et~al.}{2012}]{angulo12}
{Angulo} R.~E.,  {Springel} V.,  {White} S.~D.~M.,  {Jenkins} A.,  {Baugh} C.~M.,   {Frenk} C.~S.,  2012, \mn@doi [\mnras] {10.1111/j.1365-2966.2012.21830.x}, \href {https://ui.adsabs.harvard.edu/abs/2012MNRAS.426.2046A} {426, 2046}

\bibitem[\protect\citeauthoryear{{Angulo}, {Zennaro}, {Contreras}, {Aric{\`o}}, {Pellejero-Iba{\~n}ez}  \& {St{\"u}cker}}{{Angulo} et~al.}{2021}]{angulo21}
{Angulo} R.~E.,  {Zennaro} M.,  {Contreras} S.,  {Aric{\`o}} G.,  {Pellejero-Iba{\~n}ez} M.,   {St{\"u}cker} J.,  2021, \mn@doi [\mnras] {10.1093/mnras/stab2018}, \href {https://ui.adsabs.harvard.edu/abs/2021MNRAS.507.5869A} {507, 5869}

\bibitem[\protect\citeauthoryear{{Aric{\`o}}, {Angulo}, {Hern{\'a}ndez-Monteagudo}, {Contreras}, {Zennaro}, {Pellejero-Iba{\~n}ez}  \& {Rosas-Guevara}}{{Aric{\`o}} et~al.}{2020}]{arico20}
{Aric{\`o}} G.,  {Angulo} R.~E.,  {Hern{\'a}ndez-Monteagudo} C.,  {Contreras} S.,  {Zennaro} M.,  {Pellejero-Iba{\~n}ez} M.,   {Rosas-Guevara} Y.,  2020, \mn@doi [\mnras] {10.1093/mnras/staa1478}, \href {https://ui.adsabs.harvard.edu/abs/2020MNRAS.495.4800A} {495, 4800}

\bibitem[\protect\citeauthoryear{{Aric{\`o}}, {Angulo}, {Hern{\'a}ndez-Monteagudo}, {Contreras}  \& {Zennaro}}{{Aric{\`o}} et~al.}{2021a}]{arico21a}
{Aric{\`o}} G.,  {Angulo} R.~E.,  {Hern{\'a}ndez-Monteagudo} C.,  {Contreras} S.,   {Zennaro} M.,  2021a, \mn@doi [\mnras] {10.1093/mnras/stab699}, \href {https://ui.adsabs.harvard.edu/abs/2021MNRAS.503.3596A} {503, 3596}

\bibitem[\protect\citeauthoryear{{Aric{\`o}}, {Angulo}, {Contreras}, {Ondaro-Mallea}, {Pellejero-Iba{\~n}ez}  \& {Zennaro}}{{Aric{\`o}} et~al.}{2021b}]{arico21b}
{Aric{\`o}} G.,  {Angulo} R.~E.,  {Contreras} S.,  {Ondaro-Mallea} L.,  {Pellejero-Iba{\~n}ez} M.,   {Zennaro} M.,  2021b, \mn@doi [\mnras] {10.1093/mnras/stab1911}, \href {https://ui.adsabs.harvard.edu/abs/2021MNRAS.506.4070A} {506, 4070}

\bibitem[\protect\citeauthoryear{{Aric{\`o}}, {Angulo}, {Zennaro}, {Contreras}, {Chen}  \& {Hern{\'a}ndez-Monteagudo}}{{Aric{\`o}} et~al.}{2023}]{arico23}
{Aric{\`o}} G.,  {Angulo} R.~E.,  {Zennaro} M.,  {Contreras} S.,  {Chen} A.,   {Hern{\'a}ndez-Monteagudo} C.,  2023, arXiv e-prints, \href {https://ui.adsabs.harvard.edu/abs/2023arXiv230305537A} {p. arXiv:2303.05537}

\bibitem[\protect\citeauthoryear{{Behroozi}, {Wechsler}  \& {Conroy}}{{Behroozi} et~al.}{2013}]{Behroozi2013}
{Behroozi} P.~S.,  {Wechsler} R.~H.,   {Conroy} C.,  2013, \mn@doi [\apj] {10.1088/0004-637X/770/1/57}, \href {http://cdsads.u-strasbg.fr/abs/2013ApJ...770...57B} {770, 57}

\bibitem[\protect\citeauthoryear{Bocquet \& Carter}{Bocquet \& Carter}{2016}]{Bocquet2016}
Bocquet S.,  Carter F.~W.,  2016, \mn@doi [The Journal of Open Source Software] {10.21105/joss.00046}, 1

\bibitem[\protect\citeauthoryear{{Bocquet} et~al.,}{{Bocquet} et~al.}{2015}]{bocquet15}
{Bocquet} S.,  et~al., 2015, \mn@doi [\apj] {10.1088/0004-637X/799/2/214}, \href {https://ui.adsabs.harvard.edu/abs/2015ApJ...799..214B} {799, 214}

\bibitem[\protect\citeauthoryear{{Bocquet} et~al.,}{{Bocquet} et~al.}{2019}]{bocquet19}
{Bocquet} S.,  et~al., 2019, \mn@doi [\apj] {10.3847/1538-4357/ab1f10}, \href {https://ui.adsabs.harvard.edu/abs/2019ApJ...878...55B} {878, 55}

\bibitem[\protect\citeauthoryear{{Bulbul} et~al.,}{{Bulbul} et~al.}{2019}]{bulbul19}
{Bulbul} E.,  et~al., 2019, \mn@doi [\apj] {10.3847/1538-4357/aaf230}, \href {https://ui.adsabs.harvard.edu/abs/2019ApJ...871...50B} {871, 50}

\bibitem[\protect\citeauthoryear{{Capasso} et~al.,}{{Capasso} et~al.}{2019}]{capasso19}
{Capasso} R.,  et~al., 2019, \mn@doi [\mnras] {10.1093/mnras/sty2645}, \href {https://ui.adsabs.harvard.edu/abs/2019MNRAS.482.1043C} {482, 1043}

\bibitem[\protect\citeauthoryear{{Carlstrom}, {Holder}  \& {Reese}}{{Carlstrom} et~al.}{2002}]{Carlstrom02}
{Carlstrom} J.~E.,  {Holder} G.~P.,   {Reese} E.~D.,  2002, \mn@doi [\araa] {10.1146/annurev.astro.40.060401.093803}, \href {https://ui.adsabs.harvard.edu/abs/2002ARA&A..40..643C} {40, 643}

\bibitem[\protect\citeauthoryear{{Chaves-Montero}, {Hern{\'a}ndez-Monteagudo}, {Angulo}  \& {Emberson}}{{Chaves-Montero} et~al.}{2021}]{chaves-montero21}
{Chaves-Montero} J.,  {Hern{\'a}ndez-Monteagudo} C.,  {Angulo} R.~E.,   {Emberson} J.~D.,  2021, \mn@doi [\mnras] {10.1093/mnras/staa3782}, \href {https://ui.adsabs.harvard.edu/abs/2021MNRAS.503.1798C} {503, 1798}

\bibitem[\protect\citeauthoryear{{Chen}, {Zhang}  \& {Yang}}{{Chen} et~al.}{2022a}]{desi-groups}
{Chen} Z.,  {Zhang} P.,   {Yang} X.,  2022a, \mn@doi [arXiv e-prints] {10.48550/arXiv.2201.12591}, \href {https://ui.adsabs.harvard.edu/abs/2022arXiv220112591C} {p. arXiv:2201.12591}

\bibitem[\protect\citeauthoryear{{Chen} et~al.,}{{Chen} et~al.}{2022b}]{chen+22}
{Chen} A.,  et~al., 2022b, arXiv e-prints, \href {https://ui.adsabs.harvard.edu/abs/2022arXiv220608591C} {p. arXiv:2206.08591}

\bibitem[\protect\citeauthoryear{{Chisari} et~al.,}{{Chisari} et~al.}{2018}]{chisari18}
{Chisari} N.~E.,  et~al., 2018, \mn@doi [\mnras] {10.1093/mnras/sty2093}, \href {https://ui.adsabs.harvard.edu/abs/2018MNRAS.480.3962C} {480, 3962}

\bibitem[\protect\citeauthoryear{{Chisari} et~al.,}{{Chisari} et~al.}{2019}]{chisari19}
{Chisari} N.~E.,  et~al., 2019, \mn@doi [The Open Journal of Astrophysics] {10.21105/astro.1905.06082}, \href {https://ui.adsabs.harvard.edu/abs/2019OJAp....2E...4C} {2, 4}

\bibitem[\protect\citeauthoryear{{Chiu} et~al.,}{{Chiu} et~al.}{2018}]{chiu18}
{Chiu} I.,  et~al., 2018, \mn@doi [\mnras] {10.1093/mnras/sty1284}, \href {https://ui.adsabs.harvard.edu/abs/2018MNRAS.478.3072C} {478, 3072}

\bibitem[\protect\citeauthoryear{{Chiu} et~al.,}{{Chiu} et~al.}{2022}]{chiu22a}
{Chiu} I.~N.,  et~al., 2022, \mn@doi [\aap] {10.1051/0004-6361/202141755}, \href {https://ui.adsabs.harvard.edu/abs/2022A&A...661A..11C} {661, A11}

\bibitem[\protect\citeauthoryear{{Chiu}, {Klein}, {Mohr}  \& {Bocquet}}{{Chiu} et~al.}{2023}]{chiu22cosmo}
{Chiu} I.~N.,  {Klein} M.,  {Mohr} J.,   {Bocquet} S.,  2023, \mn@doi [\mnras] {10.1093/mnras/stad957}, \href {https://ui.adsabs.harvard.edu/abs/2023MNRAS.522.1601C} {522, 1601}

\bibitem[\protect\citeauthoryear{{Contreras}, {Angulo}, {Zennaro}, {Aric{\`o}}  \& {Pellejero-Iba{\~n}ez}}{{Contreras} et~al.}{2020}]{contreras20}
{Contreras} S.,  {Angulo} R.~E.,  {Zennaro} M.,  {Aric{\`o}} G.,   {Pellejero-Iba{\~n}ez} M.,  2020, \mn@doi [\mnras] {10.1093/mnras/staa3117}, \href {https://ui.adsabs.harvard.edu/abs/2020MNRAS.499.4905C} {499, 4905}

\bibitem[\protect\citeauthoryear{{Crain} et~al.,}{{Crain} et~al.}{2015}]{Crain2015}
{Crain} R.~A.,  et~al., 2015, \mn@doi [\mnras] {10.1093/mnras/stv725}, \href {https://ui.adsabs.harvard.edu/abs/2015MNRAS.450.1937C} {450, 1937}

\bibitem[\protect\citeauthoryear{{Croston} et~al.,}{{Croston} et~al.}{2008}]{croston08}
{Croston} J.~H.,  et~al., 2008, \mn@doi [\aap] {10.1051/0004-6361:20079154}, \href {https://ui.adsabs.harvard.edu/abs/2008A&A...487..431C} {487, 431}

\bibitem[\protect\citeauthoryear{{Dalal} et~al.,}{{Dalal} et~al.}{2023}]{Dalal2023}
{Dalal} R.,  et~al., 2023, \mn@doi [arXiv e-prints] {10.48550/arXiv.2304.00701}, \href {https://ui.adsabs.harvard.edu/abs/2023arXiv230400701D} {p. arXiv:2304.00701}

\bibitem[\protect\citeauthoryear{{Dark Energy Survey \& Kilo-Degree Survey Collaboration} et~al.,}{{Dark Energy Survey \& Kilo-Degree Survey Collaboration} et~al.}{2023}]{des+kids}
{Dark Energy Survey \& Kilo-Degree Survey Collaboration} et~al., 2023, \mn@doi [arXiv e-prints] {10.48550/arXiv.2305.17173}, \href {https://ui.adsabs.harvard.edu/abs/2023arXiv230517173E} {p. arXiv:2305.17173}

\bibitem[\protect\citeauthoryear{{Debackere}, {Schaye}  \& {Hoekstra}}{{Debackere} et~al.}{2020}]{debackere20}
{Debackere} S. N.~B.,  {Schaye} J.,   {Hoekstra} H.,  2020, \mn@doi [\mnras] {10.1093/mnras/stz3446}, \href {https://ui.adsabs.harvard.edu/abs/2020MNRAS.492.2285D} {492, 2285}

\bibitem[\protect\citeauthoryear{{Debackere}, {Schaye}  \& {Hoekstra}}{{Debackere} et~al.}{2021}]{debackere21}
{Debackere} S. N.~B.,  {Schaye} J.,   {Hoekstra} H.,  2021, \mn@doi [\mnras] {10.1093/mnras/stab1326}, \href {https://ui.adsabs.harvard.edu/abs/2021MNRAS.505..593D} {505, 593}

\bibitem[\protect\citeauthoryear{{Dietrich} et~al.,}{{Dietrich} et~al.}{2019}]{dietrich19}
{Dietrich} J.~P.,  et~al., 2019, \mn@doi [\mnras] {10.1093/mnras/sty3088}, \href {https://ui.adsabs.harvard.edu/abs/2019MNRAS.483.2871D} {483, 2871}

\bibitem[\protect\citeauthoryear{{Eckert} et~al.,}{{Eckert} et~al.}{2016}]{eckert16}
{Eckert} D.,  et~al., 2016, \mn@doi [\aap] {10.1051/0004-6361/201527293}, \href {https://ui.adsabs.harvard.edu/abs/2016A&A...592A..12E} {592, A12}

\bibitem[\protect\citeauthoryear{{Eckert}, {Ettori}, {Pointecouteau}, {Molendi}, {Paltani}  \& {Tchernin}}{{Eckert} et~al.}{2017}]{eckert17}
{Eckert} D.,  {Ettori} S.,  {Pointecouteau} E.,  {Molendi} S.,  {Paltani} S.,   {Tchernin} C.,  2017, \mn@doi [Astronomische Nachrichten] {10.1002/asna.201713345}, \href {https://ui.adsabs.harvard.edu/abs/2017AN....338..293E} {338, 293}

\bibitem[\protect\citeauthoryear{{Eckert}, {Gaspari}, {Gastaldello}, {Le Brun}  \& {O'Sullivan}}{{Eckert} et~al.}{2021}]{Eckert21}
{Eckert} D.,  {Gaspari} M.,  {Gastaldello} F.,  {Le Brun} A. M.~C.,   {O'Sullivan} E.,  2021, \mn@doi [Universe] {10.3390/universe7050142}, \href {https://ui.adsabs.harvard.edu/abs/2021Univ....7..142E} {7, 142}

\bibitem[\protect\citeauthoryear{{Ettori}, {Gastaldello}, {Leccardi}, {Molendi}, {Rossetti}, {Buote}  \& {Meneghetti}}{{Ettori} et~al.}{2010}]{ettori10}
{Ettori} S.,  {Gastaldello} F.,  {Leccardi} A.,  {Molendi} S.,  {Rossetti} M.,  {Buote} D.,   {Meneghetti} M.,  2010, \mn@doi [\aap] {10.1051/0004-6361/201015271}, \href {https://ui.adsabs.harvard.edu/abs/2010A&A...524A..68E} {524, A68}

\bibitem[\protect\citeauthoryear{{Farahi} et~al.,}{{Farahi} et~al.}{2019}]{farahi19}
{Farahi} A.,  et~al., 2019, \mn@doi [Nature Communications] {10.1038/s41467-019-10471-y}, \href {https://ui.adsabs.harvard.edu/abs/2019NatCo..10.2504F} {10, 2504}

\bibitem[\protect\citeauthoryear{{Gatti} et~al.,}{{Gatti} et~al.}{2022}]{gatti22}
{Gatti} M.,  et~al., 2022, \mn@doi [\prd] {10.1103/PhysRevD.105.123525}, \href {https://ui.adsabs.harvard.edu/abs/2022PhRvD.105l3525G} {105, 123525}

\bibitem[\protect\citeauthoryear{{Ghirardini} et~al.,}{{Ghirardini} et~al.}{2019}]{ghirardini19}
{Ghirardini} V.,  et~al., 2019, \mn@doi [\aap] {10.1051/0004-6361/201833325}, \href {https://ui.adsabs.harvard.edu/abs/2019A&A...621A..41G} {621, A41}

\bibitem[\protect\citeauthoryear{{Giri} \& {Schneider}}{{Giri} \& {Schneider}}{2021}]{giri21}
{Giri} S.~K.,  {Schneider} A.,  2021, \mn@doi [\jcap] {10.1088/1475-7516/2021/12/046}, \href {https://ui.adsabs.harvard.edu/abs/2021JCAP...12..046G} {2021, 046}

\bibitem[\protect\citeauthoryear{{Goodman} \& {Weare}}{{Goodman} \& {Weare}}{2010}]{emcee}
{Goodman} J.,  {Weare} J.,  2010, \mn@doi [Communications in Applied Mathematics and Computational Science] {10.2140/camcos.2010.5.65}, \href {https://ui.adsabs.harvard.edu/abs/2010CAMCS...5...65G} {5, 65}

\bibitem[\protect\citeauthoryear{{Grandis} et~al.,}{{Grandis} et~al.}{2020}]{grandis20}
{Grandis} S.,  et~al., 2020, \mn@doi [\mnras] {10.1093/mnras/staa2333}, \href {https://ui.adsabs.harvard.edu/abs/2020MNRAS.498..771G} {498, 771}

\bibitem[\protect\citeauthoryear{{Grandis} et~al.,}{{Grandis} et~al.}{2021a}]{grandis21}
{Grandis} S.,  et~al., 2021a, \mn@doi [\mnras] {10.1093/mnras/stab869}, \href {https://ui.adsabs.harvard.edu/abs/2021MNRAS.504.1253G} {504, 1253}

\bibitem[\protect\citeauthoryear{{Grandis}, {Bocquet}, {Mohr}, {Klein}  \& {Dolag}}{{Grandis} et~al.}{2021b}]{grandis21b}
{Grandis} S.,  {Bocquet} S.,  {Mohr} J.~J.,  {Klein} M.,   {Dolag} K.,  2021b, \mn@doi [\mnras] {10.1093/mnras/stab2414}, \href {https://ui.adsabs.harvard.edu/abs/2021MNRAS.507.5671G} {507, 5671}

\bibitem[\protect\citeauthoryear{{Harnois-D{\'e}raps}, {van Waerbeke}, {Viola}  \& {Heymans}}{{Harnois-D{\'e}raps} et~al.}{2015}]{Harnois15}
{Harnois-D{\'e}raps} J.,  {van Waerbeke} L.,  {Viola} M.,   {Heymans} C.,  2015, \mn@doi [\mnras] {10.1093/mnras/stv646}, \href {https://ui.adsabs.harvard.edu/abs/2015MNRAS.450.1212H} {450, 1212}

\bibitem[\protect\citeauthoryear{{Huang} et~al.,}{{Huang} et~al.}{2021}]{huang21}
{Huang} H.-J.,  et~al., 2021, \mn@doi [\mnras] {10.1093/mnras/stab357}, \href {https://ui.adsabs.harvard.edu/abs/2021MNRAS.502.6010H} {502, 6010}

\bibitem[\protect\citeauthoryear{{Hurier} \& {Angulo}}{{Hurier} \& {Angulo}}{2018}]{Hurier18}
{Hurier} G.,  {Angulo} R.~E.,  2018, \mn@doi [\aap] {10.1051/0004-6361/201731999}, \href {https://ui.adsabs.harvard.edu/abs/2018A&A...610L...4H} {610, L4}

\bibitem[\protect\citeauthoryear{{Huterer}}{{Huterer}}{2022}]{huterer23}
{Huterer} D.,  2022, \mn@doi [arXiv e-prints] {10.48550/arXiv.2212.05003}, \href {https://ui.adsabs.harvard.edu/abs/2022arXiv221205003H} {p. arXiv:2212.05003}

\bibitem[\protect\citeauthoryear{{Kaiser}}{{Kaiser}}{1986}]{Kaiser86}
{Kaiser} N.,  1986, \mn@doi [\mnras] {10.1093/mnras/222.2.323}, \href {https://ui.adsabs.harvard.edu/abs/1986MNRAS.222..323K} {222, 323}

\bibitem[\protect\citeauthoryear{{Krause} et~al.,}{{Krause} et~al.}{2021}]{krause21}
{Krause} E.,  et~al., 2021, \mn@doi [arXiv e-prints] {10.48550/arXiv.2105.13548}, \href {https://ui.adsabs.harvard.edu/abs/2021arXiv210513548K} {p. arXiv:2105.13548}

\bibitem[\protect\citeauthoryear{{Kravtsov}, {Vikhlinin}  \& {Meshcheryakov}}{{Kravtsov} et~al.}{2018}]{Kravtsov2018}
{Kravtsov} A.~V.,  {Vikhlinin} A.~A.,   {Meshcheryakov} A.~V.,  2018, \mn@doi [Astronomy Letters] {10.1134/S1063773717120015}, \href {http://adsabs.harvard.edu/abs/2018AstL...44....8K} {44, 8}

\bibitem[\protect\citeauthoryear{{Kugel} et~al.,}{{Kugel} et~al.}{2023}]{flamingo2}
{Kugel} R.,  et~al., 2023, \mn@doi [arXiv e-prints] {10.48550/arXiv.2306.05492}, \href {https://ui.adsabs.harvard.edu/abs/2023arXiv230605492K} {p. arXiv:2306.05492}

\bibitem[\protect\citeauthoryear{{Li} et~al.,}{{Li} et~al.}{2023}]{Li2023}
{Li} X.,  et~al., 2023, \mn@doi [arXiv e-prints] {10.48550/arXiv.2304.00702}, \href {https://ui.adsabs.harvard.edu/abs/2023arXiv230400702L} {p. arXiv:2304.00702}

\bibitem[\protect\citeauthoryear{{Macquart} et~al.,}{{Macquart} et~al.}{2020}]{macquart20}
{Macquart} J.~P.,  et~al., 2020, \mn@doi [\nat] {10.1038/s41586-020-2300-2}, \href {https://ui.adsabs.harvard.edu/abs/2020Natur.581..391M} {581, 391}

\bibitem[\protect\citeauthoryear{{Mantz} et~al.,}{{Mantz} et~al.}{2015}]{mantz15}
{Mantz} A.~B.,  et~al., 2015, \mn@doi [\mnras] {10.1093/mnras/stu2096}, \href {https://ui.adsabs.harvard.edu/abs/2015MNRAS.446.2205M} {446, 2205}

\bibitem[\protect\citeauthoryear{{Mantz} et~al.,}{{Mantz} et~al.}{2016}]{mantz16}
{Mantz} A.~B.,  et~al., 2016, \mn@doi [\mnras] {10.1093/mnras/stw2250}, \href {https://ui.adsabs.harvard.edu/abs/2016MNRAS.463.3582M} {463, 3582}

\bibitem[\protect\citeauthoryear{{Martinet}, {Castro}, {Harnois-D{\'e}raps}, {Jullo}, {Giocoli}  \& {Dolag}}{{Martinet} et~al.}{2021}]{martinet21}
{Martinet} N.,  {Castro} T.,  {Harnois-D{\'e}raps} J.,  {Jullo} E.,  {Giocoli} C.,   {Dolag} K.,  2021, \mn@doi [\aap] {10.1051/0004-6361/202040155}, \href {https://ui.adsabs.harvard.edu/abs/2021A&A...648A.115M} {648, A115}

\bibitem[\protect\citeauthoryear{{McAlpine} et~al.,}{{McAlpine} et~al.}{2016}]{McAlpine2016}
{McAlpine} S.,  et~al., 2016, \mn@doi [Astronomy and Computing] {10.1016/j.ascom.2016.02.004}, \href {https://ui.adsabs.harvard.edu/abs/2016A&C....15...72M} {15, 72}

\bibitem[\protect\citeauthoryear{{McCarthy}, {Schaye}, {Bird}  \& {Le Brun}}{{McCarthy} et~al.}{2017}]{maccarthy17}
{McCarthy} I.~G.,  {Schaye} J.,  {Bird} S.,   {Le Brun} A. M.~C.,  2017, \mn@doi [\mnras] {10.1093/mnras/stw2792}, \href {https://ui.adsabs.harvard.edu/abs/2017MNRAS.465.2936M} {465, 2936}

\bibitem[\protect\citeauthoryear{{McCarthy}, {Bird}, {Schaye}, {Harnois-Deraps}, {Font}  \& {van Waerbeke}}{{McCarthy} et~al.}{2018}]{mccarthy18}
{McCarthy} I.~G.,  {Bird} S.,  {Schaye} J.,  {Harnois-Deraps} J.,  {Font} A.~S.,   {van Waerbeke} L.,  2018, \mn@doi [\mnras] {10.1093/mnras/sty377}, \href {https://ui.adsabs.harvard.edu/abs/2018MNRAS.476.2999M} {476, 2999}

\bibitem[\protect\citeauthoryear{{McDonald} et~al.,}{{McDonald} et~al.}{2013}]{mcdonald13}
{McDonald} M.,  et~al., 2013, \mn@doi [\apj] {10.1088/0004-637X/774/1/23}, \href {https://ui.adsabs.harvard.edu/abs/2013ApJ...774...23M} {774, 23}

\bibitem[\protect\citeauthoryear{{Mead}, {Peacock}, {Heymans}, {Joudaki}  \& {Heavens}}{{Mead} et~al.}{2015}]{mead15}
{Mead} A.~J.,  {Peacock} J.~A.,  {Heymans} C.,  {Joudaki} S.,   {Heavens} A.~F.,  2015, \mn@doi [\mnras] {10.1093/mnras/stv2036}, \href {https://ui.adsabs.harvard.edu/abs/2015MNRAS.454.1958M} {454, 1958}

\bibitem[\protect\citeauthoryear{{Mead}, {Brieden}, {Tr{\"o}ster}  \& {Heymans}}{{Mead} et~al.}{2021}]{mead21}
{Mead} A.~J.,  {Brieden} S.,  {Tr{\"o}ster} T.,   {Heymans} C.,  2021, \mn@doi [\mnras] {10.1093/mnras/stab082}, \href {https://ui.adsabs.harvard.edu/abs/2021MNRAS.502.1401M} {502, 1401}

\bibitem[\protect\citeauthoryear{{Mulroy} et~al.,}{{Mulroy} et~al.}{2019}]{Mulroy19}
{Mulroy} S.~L.,  et~al., 2019, \mn@doi [\mnras] {10.1093/mnras/sty3484}, \href {https://ui.adsabs.harvard.edu/abs/2019MNRAS.484...60M} {484, 60}

\bibitem[\protect\citeauthoryear{{Navarro}, {Frenk}  \& {White}}{{Navarro} et~al.}{1996}]{nfw}
{Navarro} J.~F.,  {Frenk} C.~S.,   {White} S. D.~M.,  1996, \mn@doi [\apj] {10.1086/177173}, \href {https://ui.adsabs.harvard.edu/abs/1996ApJ...462..563N} {462, 563}

\bibitem[\protect\citeauthoryear{{Nicola} et~al.,}{{Nicola} et~al.}{2022}]{nicola22}
{Nicola} A.,  et~al., 2022, \mn@doi [\jcap] {10.1088/1475-7516/2022/04/046}, \href {https://ui.adsabs.harvard.edu/abs/2022JCAP...04..046N} {2022, 046}

\bibitem[\protect\citeauthoryear{{Pakmor} et~al.,}{{Pakmor} et~al.}{2022}]{milleniumTNG}
{Pakmor} R.,  et~al., 2022, \mn@doi [arXiv e-prints] {10.48550/arXiv.2210.10060}, \href {https://ui.adsabs.harvard.edu/abs/2022arXiv221010060P} {p. arXiv:2210.10060}

\bibitem[\protect\citeauthoryear{{Pandey} et~al.,}{{Pandey} et~al.}{2019}]{pandey19}
{Pandey} S.,  et~al., 2019, \mn@doi [\prd] {10.1103/PhysRevD.100.063519}, \href {https://ui.adsabs.harvard.edu/abs/2019PhRvD.100f3519P} {100, 063519}

\bibitem[\protect\citeauthoryear{{Pillepich} et~al.,}{{Pillepich} et~al.}{2018}]{pillepich18}
{Pillepich} A.,  et~al., 2018, \mn@doi [\mnras] {10.1093/mnras/stx2656}, \href {https://ui.adsabs.harvard.edu/abs/2018MNRAS.473.4077P} {473, 4077}

\bibitem[\protect\citeauthoryear{{Popesso} et~al.,}{{Popesso} et~al.}{2023}]{popesso23}
{Popesso} P.,  et~al., 2023, \mn@doi [arXiv e-prints] {10.48550/arXiv.2302.08405}, \href {https://ui.adsabs.harvard.edu/abs/2023arXiv230208405P} {p. arXiv:2302.08405}

\bibitem[\protect\citeauthoryear{{Predehl} et~al.,}{{Predehl} et~al.}{2021}]{predehl21}
{Predehl} P.,  et~al., 2021, \mn@doi [\aap] {10.1051/0004-6361/202039313}, \href {https://ui.adsabs.harvard.edu/abs/2021A&A...647A...1P} {647, A1}

\bibitem[\protect\citeauthoryear{{Preston}, {Amon}  \& {Efstathiou}}{{Preston} et~al.}{2023}]{preston+23}
{Preston} C.,  {Amon} A.,   {Efstathiou} G.,  2023, \mn@doi [arXiv e-prints] {10.48550/arXiv.2305.09827}, \href {https://ui.adsabs.harvard.edu/abs/2023arXiv230509827P} {p. arXiv:2305.09827}

\bibitem[\protect\citeauthoryear{{Qu} et~al.,}{{Qu} et~al.}{2023}]{act-cmbl}
{Qu} F.~J.,  et~al., 2023, \mn@doi [arXiv e-prints] {10.48550/arXiv.2304.05202}, \href {https://ui.adsabs.harvard.edu/abs/2023arXiv230405202Q} {p. arXiv:2304.05202}

\bibitem[\protect\citeauthoryear{{Ragagnin}, {Saro}, {Singh}  \& {Dolag}}{{Ragagnin} et~al.}{2021}]{Ragagnin21}
{Ragagnin} A.,  {Saro} A.,  {Singh} P.,   {Dolag} K.,  2021, \mn@doi [\mnras] {10.1093/mnras/staa3523}, \href {https://ui.adsabs.harvard.edu/abs/2021MNRAS.500.5056R} {500, 5056}

\bibitem[\protect\citeauthoryear{{Ragagnin}, {Andreon}  \& {Puddu}}{{Ragagnin} et~al.}{2022}]{ragagnin22}
{Ragagnin} A.,  {Andreon} S.,   {Puddu} E.,  2022, \mn@doi [\aap] {10.1051/0004-6361/202244397}, \href {https://ui.adsabs.harvard.edu/abs/2022A&A...666A..22R} {666, A22}

\bibitem[\protect\citeauthoryear{{Salcido}, {McCarthy}, {Kwan}, {Upadhye}  \& {Font}}{{Salcido} et~al.}{2023}]{Antilles2023}
{Salcido} J.,  {McCarthy} I.~G.,  {Kwan} J.,  {Upadhye} A.,   {Font} A.~S.,  2023, \mn@doi [\mnras] {10.1093/mnras/stad1474}, \href {https://ui.adsabs.harvard.edu/abs/2023MNRAS.523.2247S} {523, 2247}

\bibitem[\protect\citeauthoryear{{Sanders}, {Fabian}, {Russell}  \& {Walker}}{{Sanders} et~al.}{2018}]{sanders17}
{Sanders} J.~S.,  {Fabian} A.~C.,  {Russell} H.~R.,   {Walker} S.~A.,  2018, \mn@doi [\mnras] {10.1093/mnras/stx2796}, \href {https://ui.adsabs.harvard.edu/abs/2018MNRAS.474.1065S} {474, 1065}

\bibitem[\protect\citeauthoryear{{Schaye} et~al.,}{{Schaye} et~al.}{2010}]{schaye10}
{Schaye} J.,  et~al., 2010, \mn@doi [\mnras] {10.1111/j.1365-2966.2009.16029.x}, \href {https://ui.adsabs.harvard.edu/abs/2010MNRAS.402.1536S} {402, 1536}

\bibitem[\protect\citeauthoryear{{Schaye} et~al.,}{{Schaye} et~al.}{2015a}]{Schaye2015}
{Schaye} J.,  et~al., 2015a, \mn@doi [\mnras] {10.1093/mnras/stu2058}, \href {https://ui.adsabs.harvard.edu/abs/2015MNRAS.446..521S} {446, 521}

\bibitem[\protect\citeauthoryear{{Schaye} et~al.,}{{Schaye} et~al.}{2015b}]{schaye15}
{Schaye} J.,  et~al., 2015b, \mn@doi [\mnras] {10.1093/mnras/stu2058}, \href {https://ui.adsabs.harvard.edu/abs/2015MNRAS.446..521S} {446, 521}

\bibitem[\protect\citeauthoryear{{Schaye} et~al.,}{{Schaye} et~al.}{2023}]{flamingo1}
{Schaye} J.,  et~al., 2023, \mn@doi [arXiv e-prints] {10.48550/arXiv.2306.04024}, \href {https://ui.adsabs.harvard.edu/abs/2023arXiv230604024S} {p. arXiv:2306.04024}

\bibitem[\protect\citeauthoryear{{Schneider} \& {Teyssier}}{{Schneider} \& {Teyssier}}{2015}]{schneider15}
{Schneider} A.,  {Teyssier} R.,  2015, \mn@doi [\jcap] {10.1088/1475-7516/2015/12/049}, \href {https://ui.adsabs.harvard.edu/abs/2015JCAP...12..049S} {2015, 049}

\bibitem[\protect\citeauthoryear{{Schneider}, {Teyssier}, {Stadel}, {Chisari}, {Le Brun}, {Amara}  \& {Refregier}}{{Schneider} et~al.}{2019}]{schneider19}
{Schneider} A.,  {Teyssier} R.,  {Stadel} J.,  {Chisari} N.~E.,  {Le Brun} A. M.~C.,  {Amara} A.,   {Refregier} A.,  2019, \mn@doi [\jcap] {10.1088/1475-7516/2019/03/020}, \href {https://ui.adsabs.harvard.edu/abs/2019JCAP...03..020S} {2019, 020}

\bibitem[\protect\citeauthoryear{{Schneider} et~al.,}{{Schneider} et~al.}{2020}]{schneider20}
{Schneider} A.,  et~al., 2020, \mn@doi [\jcap] {10.1088/1475-7516/2020/04/020}, \href {https://ui.adsabs.harvard.edu/abs/2020JCAP...04..020S} {2020, 020}

\bibitem[\protect\citeauthoryear{{Schneider}, {Giri}, {Amodeo}  \& {Refregier}}{{Schneider} et~al.}{2022}]{schnieder21}
{Schneider} A.,  {Giri} S.~K.,  {Amodeo} S.,   {Refregier} A.,  2022, \mn@doi [\mnras] {10.1093/mnras/stac1493}, \href {https://ui.adsabs.harvard.edu/abs/2022MNRAS.514.3802S} {514, 3802}

\bibitem[\protect\citeauthoryear{{Schrabback} et~al.,}{{Schrabback} et~al.}{2021}]{Schrabback21}
{Schrabback} T.,  et~al., 2021, \mn@doi [\mnras] {10.1093/mnras/stab1386}, \href {https://ui.adsabs.harvard.edu/abs/2021MNRAS.505.3923S} {505, 3923}

\bibitem[\protect\citeauthoryear{{Sereno} \& {Ettori}}{{Sereno} \& {Ettori}}{2017}]{sereno17}
{Sereno} M.,  {Ettori} S.,  2017, \mn@doi [\mnras] {10.1093/mnras/stx576}, \href {https://ui.adsabs.harvard.edu/abs/2017MNRAS.468.3322S} {468, 3322}

\bibitem[\protect\citeauthoryear{{Springel} et~al.,}{{Springel} et~al.}{2018}]{springel18}
{Springel} V.,  et~al., 2018, \mn@doi [\mnras] {10.1093/mnras/stx3304}, \href {https://ui.adsabs.harvard.edu/abs/2018MNRAS.475..676S} {475, 676}

\bibitem[\protect\citeauthoryear{{Stern} et~al.,}{{Stern} et~al.}{2019}]{stern19}
{Stern} C.,  et~al., 2019, \mn@doi [\mnras] {10.1093/mnras/stz234}, \href {https://ui.adsabs.harvard.edu/abs/2019MNRAS.485...69S} {485, 69}

\bibitem[\protect\citeauthoryear{{Tr{\"o}ster} et~al.,}{{Tr{\"o}ster} et~al.}{2022}]{troester22}
{Tr{\"o}ster} T.,  et~al., 2022, \mn@doi [\aap] {10.1051/0004-6361/202142197}, \href {https://ui.adsabs.harvard.edu/abs/2022A&A...660A..27T} {660, A27}

\bibitem[\protect\citeauthoryear{{Umetsu} et~al.,}{{Umetsu} et~al.}{2020}]{umetsu20}
{Umetsu} K.,  et~al., 2020, \mn@doi [\apj] {10.3847/1538-4357/ab6bca}, \href {https://ui.adsabs.harvard.edu/abs/2020ApJ...890..148U} {890, 148}

\bibitem[\protect\citeauthoryear{Villaescusa-Navarro et~al.}{Villaescusa-Navarro et~al.}{2021}]{CAMELS:2020cof}
Villaescusa-Navarro F.,  et~al., 2021, \mn@doi [Astrophys. J.] {10.3847/1538-4357/abf7ba}, 915, 71

\bibitem[\protect\citeauthoryear{{Vogelsberger} et~al.,}{{Vogelsberger} et~al.}{2014}]{vogelsberger10}
{Vogelsberger} M.,  et~al., 2014, \mn@doi [\mnras] {10.1093/mnras/stu1536}, \href {https://ui.adsabs.harvard.edu/abs/2014MNRAS.444.1518V} {444, 1518}

\bibitem[\protect\citeauthoryear{{Yoon} \& {Jee}}{{Yoon} \& {Jee}}{2021}]{yoon21}
{Yoon} M.,  {Jee} M.~J.,  2021, \mn@doi [\apj] {10.3847/1538-4357/abcd9e}, \href {https://ui.adsabs.harvard.edu/abs/2021ApJ...908...13Y} {908, 13}

\bibitem[\protect\citeauthoryear{{Zennaro}, {Angulo}, {Pellejero-Ib{\'a}{\~n}ez}, {St{\"u}cker}, {Contreras}  \& {Aric{\`o}}}{{Zennaro} et~al.}{2023}]{zennaro23}
{Zennaro} M.,  {Angulo} R.~E.,  {Pellejero-Ib{\'a}{\~n}ez} M.,  {St{\"u}cker} J.,  {Contreras} S.,   {Aric{\`o}} G.,  2023, \mn@doi [\mnras] {10.1093/mnras/stad2008}, \href {https://ui.adsabs.harvard.edu/abs/2023MNRAS.tmp.1973Z} {}

\bibitem[\protect\citeauthoryear{{de Haan} et~al.,}{{de Haan} et~al.}{2016}]{dehaan+16}
{de Haan} T.,  et~al., 2016, \mn@doi [\apj] {10.3847/0004-637X/832/1/95}, \href {https://ui.adsabs.harvard.edu/abs/2016ApJ...832...95D} {832, 95}

\bibitem[\protect\citeauthoryear{{van Daalen}, {McCarthy}  \& {Schaye}}{{van Daalen} et~al.}{2020}]{vandaalen20}
{van Daalen} M.~P.,  {McCarthy} I.~G.,   {Schaye} J.,  2020, \mn@doi [\mnras] {10.1093/mnras/stz3199}, \href {https://ui.adsabs.harvard.edu/abs/2020MNRAS.491.2424V} {491, 2424}

\makeatother
\end{thebibliography}




\appendix
\section{Conversion of gas fraction between over densities}\label{app:500c-corr}

The gas mass fraction w.r.t. a given over density $\Delta$c is given by the ratio of the gas mass $M_\text{ICM}(<r_{\Delta\text{c}})$ enclosed within the radius $r_{\Delta\text{c}}$ corresponding to that over density, and the total mass $M_{\Delta\text{c}}$ w.r.t. that over density, i.e.
\begin{equation}
    f^{\Delta\text{c}}_\text{ICM} = \frac{M_\text{ICM}(<r_{\Delta\text{c}})}{M_{\Delta\text{c}}}.
\end{equation} 

Thus, the ratio between the gas mass fractions at two different over densities $\Delta$c and $\Delta^\prime$c, can be readily expressed by the ratios of gas masses and total masses, as
\begin{equation}
    \frac{f^{\Delta\text{c}}_\text{ICM}}{f^{\Delta^\prime \text{c}}_\text{ICM}} = \frac{M_\text{ICM}(<r_{\Delta\text{c}})}{M_\text{ICM}(<r_{\Delta^\prime\text{c}})} \frac{M_{\Delta^\prime\text{c}}}{M_{\Delta\text{c}}}.
\end{equation}
The second term is computed via the customary transformations between over density masses at different over densities \citep[see for instance][App.~A]{ettori10}, assuming the concentration mass relation by \citet{Ragagnin21} at its pivot cosmology.

The ratio between the gas masses can be estimated from the bound gas profile $\rho_\text{bg}(r)$ as
\begin{equation}
    \frac{M_\text{ICM}(<r_{\Delta\text{c}})}{M_\text{ICM}(<r_{\Delta^\prime\text{c}})} = \int_0^{r_{\Delta\text{c}}} r^2\text{d}r\, \rho_\text{ICM}(r) \Big/ \int_0^{r_{\Delta^\prime\text{c}}} r^2\text{d}r\, \rho_\text{ICM}(r),
\end{equation}
which is a function of the BCM parameters.

We assume that the stellar fraction stays constant, as the brightest central galaxy is included in both $r_{500\text{c}}$ and $r_{500\text{c}}$, and the rest of the profile follows the dark matter profile.

\section{Particle Radius in \texttt{bacco}}\label{app:particle-radius}

\begin{figure}
	\includegraphics[width=\columnwidth]{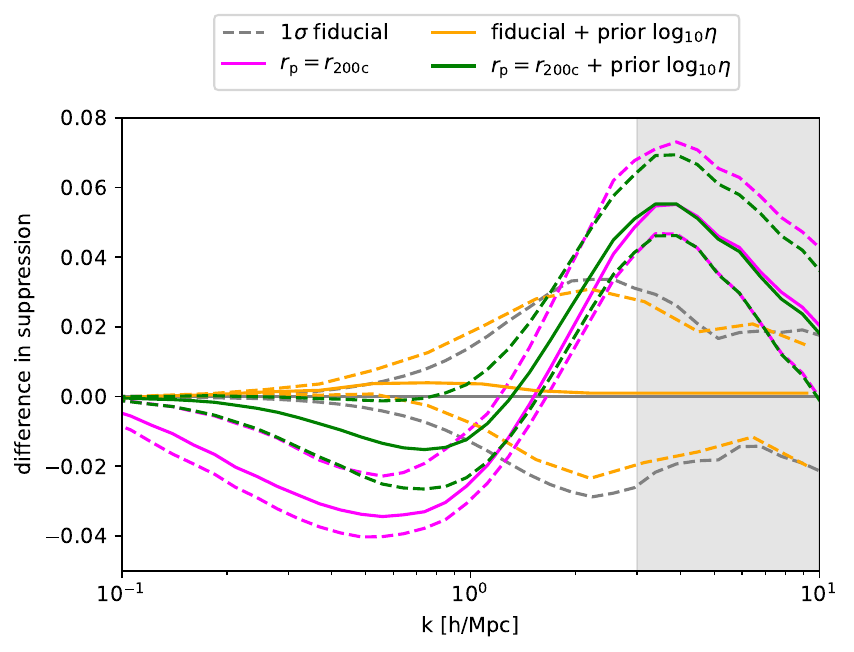}
    \caption{Differences in the matter power suppression predicted by our posterior when modifying the maximum particle displacement radius $r_\text{p}$ (fiducial: $r_\text{p}=r_\text{out}$), and when applying a hydro simulation prior of the position of the ejected gas $\eta$ in the $\texttt{bacco}$ model. Full lines represent the 50th percentile of the posterior predictive distribution at each scale, with the dashed lines represent the 16th and 84th percentile. Deviations are not larger than 2 $\sigma$, while absolute difference are of the same order as the posterior predictive precision.} 
    \label{fig:rp}
\end{figure}

One of the main differences between S19 and the \texttt{bacco} model is that the former displaces formally all the particles in the simulation (although the displacement tends to zero at large distances from halos), whereas the latter displaces only particles in haloes, inside a radius that we shall call the particle radius, $r_\text{p}$. The BCM analytical halo density profiles are opportunely truncated at $r_\text{p}$, so that a long-range displacement causes particles to be ejected from the haloes and trace the analytical ejected gas profiles \citep[see ][for more details]{arico20}. In \cite{arico20}, and following works, it was fixed $r_\text{p} = r_\text{200c}$. This choice was motivated by the fact that the parametrisation and priors used were such that, beyond $r_\text{200c}$, the ejected gas was the only component with a shape different from NFW. However, in this work, we utilize X-ray density profile to constrain $r_\text{out}>r_\text{200c}$, meaning that the hot gas does not follow the dark matter beyond $r_\text{200c}$ (we remind the reader that in \texttt{bacco} the hot gas profile follows the NFW after $r_\text{out}$). 

For this work, we decided then to set $r_\text{p}=r_\text{out}$, to enforce the particle radius to be at the exact position of the change in slope of the hot gas to NFW. The choice of $r_\text{p}$ leads to a qualitatively different -- though statistically consistent -- matter power suppression, as shown in Fig~\ref{fig:rp}, grey and magenta lines. Specifically, we find that setting $r_\text{p}=r_\text{out}$ the suppression in the matter power spectrum is reduced at large scales, and enhanced at small scales, with respect to $r_\text{p} = r_\text{200c}$.
We note that, while the free parameters of \texttt{bacco} act on the shape and normalisation of the analytical density profiles, the particle radius $r_\text{p}$ impacts directly the displacement fields applied to the particles. Smaller $r_\text{p}$ drive larger displacements fields, and therefore more suppression at large scales. e.g. in \cite{arico21b}. Vice versa, larger $r_\text{p}$ cause on average smaller displacements in the particles, and therefore the suppression in the matter power spectrum is more similar to S19.

The large scale shape of the matter power spectrum suppression predicted by $\texttt{bacco}$ is regulated by the position of the ejected gas $\eta$. X-ray observations of galaxy clusters are not able to probe the ejected gas directly, as the X-ray surface brightness is proportional to the gas density squared. They just place an upper limit on the amount of ejected gas by measuring the amount of gas retained by massive halos, without determining its location outside of the halo. We thus explore the impact of a Gaussian prior $\log_{10} \eta = -0.32 \pm 0.22$ derived from $74$ hydro-dynamical simulations \citep{arico21b}. This positions the ejected gas in a physically reasonable region. The impact of this prior on the matter power spectrum suppression is also shown in Fig.~\ref{fig:rp}, orange and green lines. In our fiducial setting, the effect is less pronounced that when choosing smaller particle radii.

In summary, the accuracy of our prediction is of the same order as its precision, that is order of a few percent. This highlights the importance of further refining and calibrating the BCMs. More detailed comparisons against hydrodynamical simulations, as well as consistency checks between different observables such as cosmic shear and X-ray data, are clearly warranted. We plan to undertake these further studies in the future, motivated by the excellent constraining power of cluster data demonstrated in this work.

\section{Marginal Plots of the Posteriors}

\begin{figure*}
	\includegraphics[width=\textwidth]{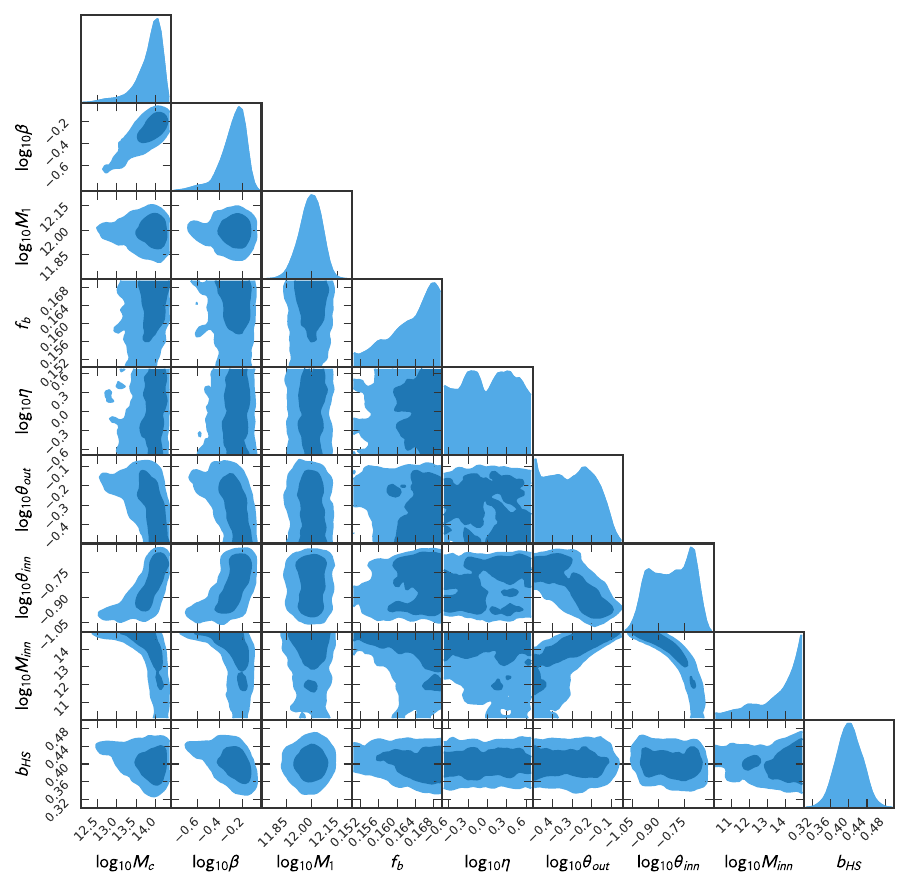}
    \caption{1-d and 2-d marginalized posterior samples resulting from the fit of the \texttt{bacco} model. The parameters $\log_{10} M_\text{c} h / M_\odot$, $ \log_{10}\beta$ and $\log_{10} M_1 h / M_\odot$ are well constrained (for the sake of a compact representation, we omit the mass units in the plot labels). We place upper limits on $\log_{10} \theta_\text{out} $ and  $\log_{10} \theta_\text{inn}$, while the other parameters remain unconstrained, with the exception of the hydrostatic mass bias, on which we placed a prior.}
    \label{fig:full_post_bacco}
\end{figure*}

\begin{figure*}
	\includegraphics[width=\textwidth]{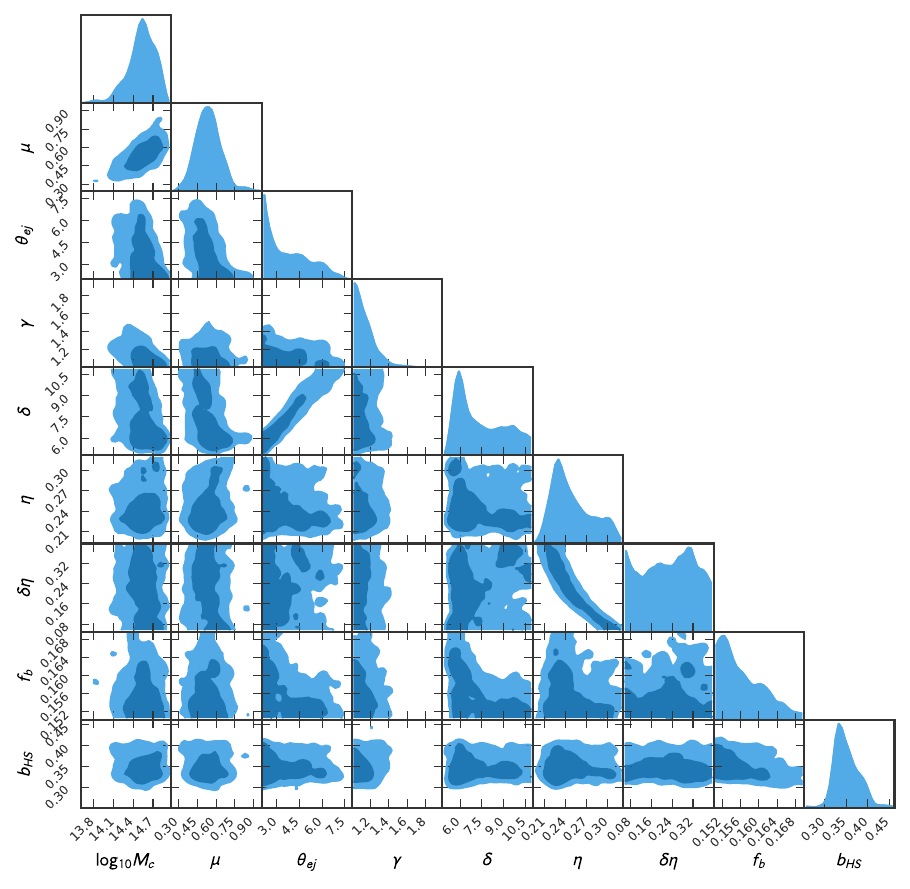}
    \caption{1-d and 2-d marginalized posterior samples resulting from the fit of the S19-model. The parameters $\log_{10} M_\text{c} h / M_\odot$, and $\mu$ are well constrained  (for the sake of a compact representation, we omit the mass units in the plot labels). We place upper limits on $\theta_\text{ej} $ and  $\gamma$, and a lower limit on $\delta$. A specific combination of the parameters $\eta$ and $\delta\eta$ is also well constrained.}
    \label{fig:full_post_AS}
\end{figure*}

In Fig.~\ref{fig:full_post_bacco} and Fig.~\ref{fig:full_post_AS} we show the 1-d and 2- marginalized posterior samples resulting from the fit of the \texttt{bacco}, and S19 model, respectively. The interpretation of these plots is provided in section~\ref{sec:posterior&gof}.

\bsp	
\label{lastpage}
\end{document}